\documentclass[10pt]{article}  
\usepackage{latexsym} 
\usepackage{amsmath}  
\begin{document} 
 
\begin{flushright} ULB--TH--00/15 \\  UMH--MG--00/04 \\July 2000\\ \end{flushright} 
\vspace{.1cm} 
\begin{center} {\large Holonomies, anomalies 
and the Fefferman-Graham ambiguity\\  
\vspace{.1cm} in $\rm AdS_3$ gravity}\\ 
\vspace{.5cm} M.~Rooman${}^{a,}$\footnote{ 
E-mail : mrooman@ulb.ac.be} and Ph.~Spindel${}^{b,}$\footnote{ E-mail : 
spindel@umh.ac.be}\\ 
\vspace{.3cm}  {${}^a${\it Service de Physique Th\'eorique}}\\  
{\it Universit\'e 
Libre de Bruxelles, Campus Plaine, C.P.225}\\ {\it Boulevard du Triomphe, 

B-1050 Bruxelles, Belgium}\\  
\vspace{.2cm} {${}^b${\it 
M\'ecanique et Gravitation}}\\ {\it Universit\'e de Mons-Hainaut, 20 
Place du Parc}\\  
{\it 7000 Mons, Belgium}\\ \end{center} 
\vspace{.1cm} 
 
\newcommand{\asym}{\mathop {\simeq} \limits_{\bar r \rightarrow \infty}} 
\newcommand{\asymr}{\mathop {\simeq} \limits_{ r \rightarrow \infty}} 
\newcommand{\real}{{\hbox{{\rm I}\kern-.2em\hbox{\rm R}}}} 
 
\begin{abstract} 
Using the Chern-Simon formulation of (2+1) gravity,  
we derive, for the general asymptotic metrics given by the 
Fefferman-Graham-Lee theorems, the emergence of the Liouville mode  
associated to the boundary degrees of freedom of (2+1) dimensional  
anti de Sitter geometries. Holonomies are
described through multi-valued gauge and Liouville fields and
are found to algebraically couple the fields defined on
the disconnected components of spatial
infinity. In the case of flat boundary metrics, explicit 
expressions are obtained for the fields and holonomies.
We also show the link between the variation 
under diffeomorphisms of the Einstein theory of gravitation and 
the Weyl anomaly of the conformal theory at infinity.
\end{abstract}
\noindent
{\it PACS\/: 11.10.Kk, 04.20.Ha }\\
{\it Keywords}: anti de Sitter, Chern-Simon, Liouville
\section{Introduction} 
 
The interest of studying (2+1) dimensional gravity has initially been  
emphasized in \cite{DJH} and have recently been revived with the discovery  
of black holes in spaces with negative cosmological constant \cite{BTZ}. 
Since then, a large number of studies has been devoted to the elucidation 
of classical as well as to quantum (2+1) gravity \cite{CAR}. The major problem 
resides in the quest for the origin of the black hole entropy. This entropy  
seems due to a macroscopically large number of physical degrees of freedom, 
as  
indicated by the value of the central charge first computed by \cite{BH}. 
 
In $(2+1)$ dimensions, the field equations of gravity 
with negative cosmological constant have been proven to be equivalent  
to those of a Chern-Simons  
(CS) theory with $SL(2,\real) \times SL(2,\real)$ as gauge group \cite{ATW}. 
Assuming the boundary of the space to be a flat cylinder $\real \times S^1$,  
Coussaert, Henneaux and van Driel (CHD) \cite{CHD} demonstrated  
the equivalence  
between this CS theory and a non-chiral Wess-Zumino-Witten (WZW) 
theory \cite{MS,EMSS}, and showed that the $\rm AdS_3$ boundary conditions as defined  
in \cite{BH} implement the constraints 
that reduce the WZW model to the Liouville theory \cite{FWB}. 
  
In this paper, we show that, using the less restrictive 
AdS boundary conditions allowed by the Fefferman-Graham-Lee 
theorems \cite{GL,FG}, the CHD analysis can be extended and leads 
to the Liouville theory formulated on a 2-dimensional curved background (a 
preliminary version of this analysis can be found in \cite{RS}). 

In section {\bf 2} we recall
the Fefferman-Graham (FG) asymptotic expansion of Einstein
and  $\rm AdS_3$ metrics. Section {\bf 3} contains the main calculation of
the paper, consisting of a generalization of the relation between
3-d gravity and Liouville theory to curved boundary metrics. Special attention
is paid to the contributions from holonomies, when the topology of the
space is cylindrical. An originality of our approach 
resides in the use of  multi-valued gauge group elements,
which allows to simply show
the correspondence between the Einstein-Hilbert (EH) 
action and a
multi-valued Liouville field action.
 
In section {\bf 4}, we discuss the coupling between the fields living on the
different components of spatial infinity, which is induced by
the holonomies. We show that this coupling 
reduces to simple algebraic conditions
on-shell. The aim of section {\bf 5} is to illustrate the link between 
Liouville fields, holonomies, $\rm AdS_3$ metrics and $SL(2,\real)$ parallel
transport matrices in the framework of flat asymptotic boundaries.
 
Finally, in section {\bf 6}, by taking into account all  
terms resulting from integrations by parts during the reduction process from 
the EH action to the Liouville action, performed in section {\bf 3}, we make 
explicitly the  
connection between the variance of the EH action under diffeomorphisms 
and Weyl transformations on the boundary of  $\rm AdS_3$ space. 
 
\section{Asymptotically anti de Sitter spaces} 
 
Graham and Lee \cite{GL} proved that, under suitable topological 
assumptions, Euclidean 
Einstein spaces with negative cosmological constant $\Lambda$ 
are completely defined by the geometry on 
their boundary. Furthermore, Fefferman and Graham (FG) \cite{FG}  
showed that, whatever the signature, 
there exists an asymptotic expansion of the metric, which formally  
solves the Einstein equations with $\Lambda<0$. 
The first terms of this expansion may be given by even powers of  
a radial coordinate $r$: 
\begin{equation} 
\label{FG} 
ds^2  
\asym 
 \ell^2 {dr^2\over{r^2}} \, + \, 
\frac {r^2} {\ell^2} \stackrel {(0)} {\bf g} (x^i)  \, + \, 
\stackrel {(2)} {\bf g} (x^i) \, + \, \cdots \qquad . 
\end{equation} 
On $(n+1)$-dimensional space-times, the full asymptotic expansion 
continues with terms of negative even  powers of $r$ up 
to $r^{-2([\frac {n+1} 2]-2)}$, with in addition a logarithmic term of  
the order 
of $r^{-(n-2)}\log r$ when $n$ is even and larger than 2. All these terms 
are completely defined by the boundary geometry  
$\stackrel {(0)} {\bf g} (x^i)$. It seems thus natural 
to take them as the definition of asymptotic Einstein metrics. 
These terms are 
followed by terms of negative powers 
starting from $r^{-2([\frac {n+1} 2]-1)}$; for even $n$, 
the trace-free part of the $r^{-(n-2)}$ 
coefficient is not fully determined by $\stackrel {(0)} {\bf g}$. It 
contains degrees of freedom in Lorentzian spaces \cite{NB,BERS}.  
Once this ambiguity is fixed,  
all subsequent terms become determined. 
 
It is instructive to look at the first iterations of this 
expansion. We write 
the metric in terms of forms $\underline{\Theta}^{ \mu}$ as 
$ds^2 = \underline{\Theta}^{ 0} \otimes \underline{\Theta}^{ 0} + 
\eta_{ a b} \, 
\underline{\Theta}^{ a} \otimes \underline{\Theta}^{ b}$ 
with $\mu, \nu$ [resp. $a, b$] running from 0 to $n$ [resp. 1 to $n$] and $\eta _{ a b}$ a 
flat $n$-dimensional minkowskian metric {\it diag.}(1,..., 1, -1). The forms  
$\underline{\Theta}^{ \mu}$ and the Levi-Civita connection 
$\underline{\Omega}^{\mu \nu}$ 
read as\footnote{By definition, $o(r^{-n})=u_n(r)$ means that 
  $\lim_{r\rightarrow \infty} \frac {u_n(r)}{r^{n}}=0$}: 
\begin{eqnarray} 
\label{FGTH} 
\underline{\Theta}^{ 0} = \ell{dr\over r} \quad &,& \quad 
\underline{\Theta}^{ a} = 
{r\over{\ell}} \underline{\theta}^{ a} + {\ell\over r} 
\underline{\sigma}^{ a} + 
o (r^{-2}) \qquad , \\ 
\label{FGOM} 
\underline{\Omega}^{a0}={r\over{\ell^2}} \underline{\theta}^{ a}  
 - {1 \over r} \underline{\tilde \sigma}^{a} + o (r^{-2}) 
 \quad &,& \quad 
\underline{\Omega}^{ab}= \underline{\omega}^{ab} +  o (r^{-1})\qquad , 
\end{eqnarray} 
where the forms $\underline {\theta}^{ a}$,  
$\underline{\omega}^{ab} \equiv {\omega}^{ab}_{\ \ c} \,  
\underline{\theta}^{ c}$, 
$\underline {\sigma}^{ a} \equiv  {\sigma}^{ a}_{\  b}  
\, \underline{\theta}^{ b}$ and 
$\underline {\tilde \sigma}^{ a} \equiv  {\sigma}_{ b}^{\  a}  
\, \underline{\theta}^{ b}$ 
are r-independent. These provide the dominant  
and sub-dominant terms of the metric expansion: 
\begin{equation} 
\label{gasymp} 
\stackrel {(0)}{\bf g}= \eta_{a   b}\, \underline{\theta}^{ a} \otimes 
\underline{\theta}^{ b} 
\quad , \quad  
\stackrel {(2)}{\bf g} \equiv 
\stackrel {(2)} g _{a b} \underline{\theta}^{ a}\otimes 
 \underline{\theta}^{ b} 
= \left ( \sigma_{ a  b} +  \sigma_{ b  a}\right ) 
\underline{\theta}^{ a} \otimes \underline{\theta}^{ b} \qquad . 
\end{equation} 
Here and in what follows, the $n$-dimensional indices 
and the covariant derivatives are defined 
with respect to the metric $\stackrel {(0)}{\bf g}$. 
Using these definitions, the $(n+1)$-dimensional 
Riemann curvature 2-form  
$\underline{\underline R}_{\mu \nu}$ becomes: 
\begin{eqnarray}   
\label{Ra0} 
\underline{\underline R}_{ a 0} &=& 
- \frac 1 {\ell^2} \underline{\Theta} _{ a} \wedge \underline{\Theta}_{ 
0}-{1\over r} 
\left(d\underline{\gamma}_{ a} + \underline{\omega}_{ a b} 
\wedge \underline{\gamma}^{ b} 
\right) + o(r^{-2}) \qquad , \\ 
\label{Rab} 
\underline{\underline R}_{ a b} &=& 
 - \frac 1 {\ell^2} \underline{\Theta}_{ a} \wedge 
\underline{\Theta}_{ b}+ 
\stackrel {(0)} {\underline{\underline {\cal R}}} _{ a b} + 
\frac 1 {\ell^2} (\underline{\theta}_{ a}\wedge \underline{\gamma}_{ b} 
+ \underline{\gamma}_{ a} \wedge \underline{\theta}_{ b}) + o(r^{-1}) \qquad , 
\end{eqnarray} 
where $ \stackrel {(0)} {\underline{\underline {\cal R}}} _{ a b}$  
is the $n$-dimensional curvature 2-form 
defined by the metric $\stackrel {(0)}{\bf g}$, 
and $\underline{\gamma}_{ a} 
\equiv \stackrel {(2)} g _{ a  b}\underline{\theta}^{ b}$. 
If we impose the metric of the $(n+1)$-dimensional space 
to be asymptotically Einsteinian, i.e.  
$R_{\mu\,\nu}=\frac 2{n-1}\Lambda\, \eta_{\mu\,\nu}$,  
these equations, at order $r^2$, fix: 
\begin{equation} 
\Lambda = -1/\ell^2\qquad.\label{Lambda} 
\end{equation}  
Moreover, at order 1 and $r^{-1}$, they yield: 
\begin{eqnarray} 
\label{Exp1} 
\stackrel {(0)} {\cal{R}}_{ a  b} + {1\over{\ell^2}} [(n-2) 
\stackrel {(2)}g_{  a   b} +  
\eta _{  a   b} \stackrel {(2)}g \strut _{ c}^{ c}] & = & 0 \qquad ,\\ 
\label{Exp2} 
 \stackrel {(2)} g \strut  ^{ b}_{ b ; a} -  
 \stackrel {(2)} g \strut ^{ b}_{ a; b} & = & 0 \qquad , 
\end{eqnarray} 
where $\stackrel {(0)}{\cal R}_{ a  b}$ are the components of the $n$-dimensional Ricci  
tensor. 
 
We define asymptotically AdS spaces by the stronger condition that the Riemann 
tensor tends to that of AdS, i.e. $\underline{\underline R}_{\mu\,\nu} = 
 \Lambda\, \underline{\Theta}_{ \mu} \wedge 
\underline{\Theta}_{\nu}$. This implies, in addition to eq. (\ref{Lambda}): 
\begin{eqnarray} 
& &\stackrel {(0)} {\cal{R}}_{ a  b\,cd} =\frac1{\ell^2}\left( 
\eta_{ad}\stackrel {(2)}g_{  bc}-\eta_{ac}\stackrel {(2)}g_{  bd}+ 
\eta_{bd}\stackrel {(2)}g_{  ac}-\eta_{bc}\stackrel {(2)}g_{  ad} 
\right)\qquad ,\label{as1}\\ 
& &\stackrel {(2)} g_{a b ; c} -  
 \stackrel {(2)} g_{ ac; b}  =  0 \qquad ,\label{as2} 
\end{eqnarray} 
whose traces are the asymptotic Einstein conditions (\ref{Exp1}, \ref{Exp2}). 
When $n \neq 2$, 
eq. (\ref{Exp1}) fully specifies the metric $\stackrel {(2)}{\bf g}$  
in terms of $\stackrel {(0)}{\bf g}$ and  
eq. (\ref{Exp2}) 
becomes the Bianchi identity satisfied by the $n$-dimensional Einstein  
tensor. Furthermore eq. (\ref{as1}) is an identity if $n=3$ and implies that 
 the Weyl tensor of the $n$-dimensional geometry vanishes if $n>3$, while 
 eq. (\ref{as2}) implies that the Cotton-York tensor vanishes for $n=3$ and 
 becomes a consequence of eq. (\ref{as1}) and the Bianchi identities for 
 $n>3$. Thus, AdS asymptotic spaces are Einstein asymptotic spaces whose 
 $\stackrel {(0)}{\bf g}$  metric tensor  is conformally flat. 
The same conclusion holds for $n=2$ as in three dimensions Einstein spaces 
 with $\Lambda <0$ are  
locally AdS and metrics on cylindrical boundaries are conformally flat. 
 
On the other hand, when $n=2$ only the  
 trace of $\stackrel {(2)}{\bf g}$  
is fixed by eq. (\ref{Exp1}): 
\begin{equation} 
\label{sigR} 
\stackrel {(2)}{g} \strut ^{ c}_{  c}= 2\, 
\sigma^{  c}_{  c} \equiv 2 \, \sigma = -{\ell^2\over 2}\stackrel {(0)}{\cal R} 
\qquad , 
\end{equation} 
and the other components of $\stackrel {(2)}{\bf g}$ have only 
to satisfy the equations: 
\begin{equation} 
\label{divg2} 
\stackrel {(2)}g \strut ^{ a}_{ b \, ;  a} = -{\ell^2\over 2}  
\stackrel {(0)} {\cal R} _{, b } 
\qquad . 
\end{equation} 
The subdominant metric components are thus not all determined by  
the asymptotic metric in three dimensions, but there remains one 
degree of freedom, which we shall explicit in the next section. 
Similar indeterminacy, involving $\stackrel {(n)}{\bf g}$, arises 
for all even values of $n$ \cite{NB, BERS}. 
 
To illustrate the meaning of the FG coordinates,  
let us consider the special case of BTZ black 
holes \cite{BTZ} whose metric can locally be written as: 
\begin{equation} 
\label{BTZmet} 
ds^2 = -(\frac {\rho^2} {\ell^2} -M) dt^2 +  
\frac {d\rho ^2} {\frac {\rho^2} {\ell^2} -M + \frac {J^2}{ 4 \rho^2}} 
+ \rho^2 d\varphi^2 + J d\varphi dt \qquad , 
\end{equation} 
where M is the mass and J the angular momentum. To shorten the discussion,  
we restrict ourselves to 
the cases $M>|J/\ell|>0$. In this case, a single coordinate patch 
$(\rho, \varphi, t)$, on which the metric is given by (\ref{BTZmet}), covers  
the three 
regions denoted I, II and III (see Fig. 1), corresponding to $\rho>\rho_+$, 
$\rho_+>\rho>\rho_-$ and $\rho<\rho_-$, where  
\begin{equation} 
\rho_+=\frac \ell 2(M +\sqrt{M^2-\frac {J^2}{\ell^2}})\qquad {\rm and} \qquad 
 \rho_-=\frac \ell 2(M -\sqrt{M^2-\frac {J^2}{\ell^2}})\qquad  
\end{equation} 
are the values of the constant $\rho$ surfaces defining the inner  
and outer horizons. The complete space-time is obtained by glueing  
similar overlapping coordinate patches. 
 
The FG radial coordinate $r$ is obtained from $\rho$ 
by the transformations: 
\begin{equation} 
\label{rrho} 
r^2  = \frac 12 \left( \rho^2 - \frac {M\ell^2} 2 \pm  
\, \ell \, \rho \, \sqrt{\frac {\rho^2}{\ell^2} -M  +  
\frac{J^2} {4 \rho^2} } \right )  \qquad , 
\end{equation} 
yielding the metric expression: 
\begin{eqnarray} 
\label{BTZFG} 
ds^2 &=& \ell^2 \frac {dr^2} {r^2} + 
\frac {r^2} {\ell^2}(\ell ^2 d\varphi^2 -d t^2)+ 
\frac M 2 (\ell ^2 d\varphi^2 +d t^2) + J dt d\varphi \nonumber \\ &&+ 
\frac {M^2 \ell^2 -J^2} {16 r^2} (\ell ^2 d\varphi^2 -d t^2) 
\quad . 
\end{eqnarray} 
This $r$ coordinate is only defined on 
region I of a coordinate patch $(\rho,\, \varphi, \, t)$,  
between the exterior horizon $\rho=\rho_+$ and infinity.  
It can be analytically continued so as to cover two adjacent regions  
I and I' (see Fig. 1). On the first (I), we have 
$r^2> \frac \ell 4 \sqrt{M^2-J^2/\ell^2}$, which is obtained 
by taking the plus sign in eq. (\ref{rrho}); on the other (I') 
$r^2< \frac \ell 4 \sqrt{M^2-J^2/\ell^2}$, corresponding 
to the minus sign. This $r$ coordinate accounts for the global 
nature of the black holes, where the  diffeomorphic regions I and I' are 
connected by an Einstein-Rosen bridge. Near the space-like 
infinity of region I  $(r \rightarrow \infty)$ the metric (\ref{BTZFG}) 
coincides with the FG 
asymptotic expansion (\ref{FG}), whereas on region I', where the space-like 
 asymptotic region is given by 
$r \rightarrow 0$, we recover  the FG expression of the asymptotic metric  
after the transformation 
$r^2 \rightarrow (M^2 \ell^2 -J^2) \ell^2 / (16 r^2)$. This transformation  
leaves the metric (\ref{BTZFG}) invariant, ensuring the consistency 
of the definitions of mass and angular momentum of the black hole, whatever 
the asymptotic region is. 
Of course we may define a single radial coordinate 
that goes to $\pm \infty$ on the asymptotia of regions I and I' and 
leads on both ends to the FG expression (\ref{FG}); it   
is however not related analytically to the $r$ coordinate (\ref{rrho}).  
We shall use such a coordinate (also called $r$) in section {\bf 3}. 
 
Furthermore, if we introduce the radial variable 
$y=r^{-2}$, corresponding to the original FG radial 
coordinate \cite{FG}, 
we may extend its domain of definition to  negative values, which 
correspond to  regions  III and III'. Note that, as $r$ has been taken 
to be intrinsically space-like, it cannot cover regions II, between 
the horizons, unless we turn it into a time-like coordinate.  
  
\section{From Einstein-Hilbert to Liouville action} 

In this section we perform explicitly the transformations leading from
the EH action to the Liouville action, assuming the metric to be asymptotically
AdS as defined in the previous section. Our first aim is to show
how the asymptotic conditions generalize in case of curved metrics,
 and lead to a Liouville action on curved background.
Our second aim is to keep track of all kinds of boundary contributions
that appear during the 
transformations (which renders this section rather long and detailed), 
in view of accounting for the effect of holonomies (see section {\bf 4})
and showing the precise link between the
EH action which is invariant under the combined action
of diffeomorphisms and Weyl transformations and the Liouville action 
which is not (see section {\bf 6}). 

\subsection{First step: Chern-Simon action}

The EH (2+1) gravity action with cosmological  
constant $\Lambda$, evaluated on a space-time domain $\cal V$, 
consists of the bulk term: 
\begin{equation} 
\label{SEH} 
S_{EH}=\frac 1 {16 \pi G} \int_{\cal V} \,  
   ( \, { \cal R} - 2 \, \Lambda ) 
       \, \boldsymbol{\eta} 
\qquad ,  
\end{equation} 
where $\boldsymbol{\eta}$ is the volume element 3-form on $\cal V$. 
 
Let us specify the topological structure of the domain of integration $\cal V$. 
Inspired by the BTZ black holes \cite{BTZ}, we assume it to consist 
of the product of a time interval $[t_0,\, t_1]$ with a spacelike 
section $\Sigma$ chosen to be homeomorphic to a disk of radius 
$\bar r$ or to a finite 2-dimensional space-like cylinder of length  
$2 \, \bar r$; afterwards $\bar r$ will be pushed to infinity.  
The disk will provide extremal black 
holes if we accept a conical singularity at the 
origin $r=0$, and global AdS space if not.  
The cylinder will describe  non-extremal black holes, its 
non trivial topology describing 
the Einstein-Rosen bridge. In fact, we may not 
exclude {\it a priori} 
more complicated topologies, such as several cylinders attached to 
an arbitrary compact manifold, but we shall not consider them here. 
 
We use usual cylindrical coordinates $(r,\, \varphi,\, t)$  
to parametrize the manifold $\cal V$, where $|r|$ coincides with the FG  
radial variable in a neighbourhood of $|r|=\infty$. 
Accordingly, the {\bf coordinate} 
domain of integration will be the cube $[t_0,\, t_1]\times 
[\bar r_L,\, \bar r_R]\times [0,\, 2\pi]$.  
In case $\Sigma$ is a disk, $\bar r_L = 0$, $\bar r_R = \bar r$,  
and all the considered quantities 
are periodic in the angular variable $\varphi$.  
In case $\Sigma$ is a cylinder, we choose  $\bar r_L =-\bar r$ and $\bar 
r_R=\bar r$;  
to account for the holonomies allowed by this non trivial 
topology, we may not assume {\it a priori} all quantities  
to be periodic in $\varphi$. Hence, the boundary terms obtained upon  
application of Stokes theorem will involve when $\Sigma$ is a disk  
the geometrical surfaces $t=t_0$, $t=t_1$ and $r=\bar r$, with in addition  
when $\Sigma$ is a cylinder the extra boundary  
component $r=- \bar r$ and the coordinate surfaces  
$\varphi=0$ and $\varphi=2 \, \pi$. 
 
When $\Lambda <0$, one can re-express \cite{ATW} the EH action (\ref{SEH})  
in terms of 
the two gauge fields  
${\bf A}=A_\mu {\underline \Theta} ^\mu = J_{ \mu} \, \underline {A}^{ \mu}$ 
and 
${\bf \tilde A}=\tilde A_\mu {\underline \Theta} ^\mu = J_{ \mu} \,  
\underline {\tilde A}^{ \mu}$,  
with $J_{ \mu}$ 
generators of the $sl(2,\real)$ algebra satisfying  
$Tr(J_\mu J_\nu)=\frac 1 2 \eta_{\mu\nu}$ and  
$Tr(J_\mu J_\nu J_\rho)=\frac 1 4 \epsilon_{\mu\nu\rho}$,  
with $\epsilon _{012}=1$ (see appendix for conventions).  
These fields are given in terms of the metric by: 
\begin{equation} 
\label{ATHOM} 
\underline{A}^{ \mu} = \frac 1 \ell \underline{\Theta}^{ \mu} + 
{1\over 2} \epsilon^{ \mu}\ _{ 
\nu\rho} \underline{\Omega}^{ \nu  \rho}  \qquad , \qquad 
\underline{\tilde A}^{ \mu} = - \frac 1 \ell \underline{\Theta}^{ \mu} + 
{1\over 2} \epsilon^{ \mu}\ _{ 
\nu\rho} \underline{\Omega}^{ \nu  \rho} \qquad  . 
\end{equation} 
They allow to express the action $S_{EH}$ as the 
difference of two CS actions $S_{SC}[{\bf A}]$ and $S_{SC}[{\bf \tilde A}]$  
plus a boundary term that mixes the two gauge fields: 
\begin{equation} 
\label{SCS} 
S_{EH}= S_{CS} + {\cal B}_{CS} 
\qquad , \qquad
S_{CS}=S[{\bf A}] - S[{\bf \tilde A}] \qquad , 
\end{equation} 
with 
\begin{equation} 
\label{SA} 
S[{\bf A}] = \frac {\ell} {16 \pi G} \int_{\cal V} 
 Tr({\bf A}\wedge d{\bf A} + {2\over 3} {\bf 
  A}\wedge {\bf A} \wedge {\bf A}) \qquad , 
\end{equation} 
and  
\begin{equation} 
\label{BEHCS} 
{\cal B}_{CS}= 
\frac {\ell} {16 \pi G} 
\int_{\cal V} Tr \, d ( {\bf \tilde A}\wedge {\bf A}) \qquad . 
\end{equation} 
The integral ${\cal B}_{CS}$ is composed of three terms, 
noted ${\cal B}_{CS}^{(r)}$, ${\cal B}_{CS}^{(\varphi)}$ and 
${\cal B}_{CS}^{(t)}$, coming from integrations 
over $r$, $\varphi$ and $t$, respectively. As we assume that the  
metric and thus the fields ${\bf A}$ and ${\bf \tilde A}$ are  
globally defined on $\cal V$, i.e. periodic in $\varphi$,  
${\cal B}_{CS}^{(\varphi)}$ vanishes. 
 
We now discuss the large $\bar r$ behaviour of the EH and CS actions  
(\ref{SEH}, \ref{SCS}).  
The insertion of eqs(\ref{FGTH}, \ref{FGOM}) in eq.(\ref{ATHOM})  
gives the asymptotic behaviour of the CS fields  
$\bf A$ and $\bf \tilde A$. But before we pursue our discussion, the following
clarifying remark seems necessary. The FG expression of the
metric is, in general, only locally valid. When  $\Sigma $ is a cylinder
there are two asymptotic regions and
we may always fix the asymptotic
behaviour of the fields on one of them according 
to eqs (\ref{FGTH}, \ref{ATHOM}). With such a
choice, the frames $\{\Theta^0,\,\Theta^a\}$ on the other
asymptotic region become defined up to a sign 
that depends on the continuation
of the CS fields across the whole manifold. The 
asymptotic behaviours of
$\bf A$ and $\bf \tilde A$ may indeed 
have to be exchanged (see section {\bf 5} for an
explicit example), but this has no consequence 
on the rest of our analysis.

Let us in a first stage focus on one asymptotic region.
To fix the notation, we adopt the convention
defined by eqs (\ref{FGTH}, \ref{ATHOM}), with the radial coordinate
$r$ belonging to a neigbourhood of $+\infty$. We
find convenient to write explicitly the large $r$ behaviour of the CS
fields using
the null frame ${\underline \theta}^{\pm} = {\underline \theta}^{1}  
\pm {\underline \theta}^{2}$ and its  
dual vectorial frame ${\vec e}_{\pm}= \frac 1 2 ({\vec e}_{1}  
\pm {\vec e}_{2})$, which are defined  on the surfaces $|r|=\bar r$.  
At order $r^{-1}$, we obtain: 
\begin{equation} 
\label{BC0} 
A_r \asymr \frac 1 {2  r} \left(\begin{array}{cc}  
  1   &  0 \\ 
  0   &   - 1  
  \end{array} \right) \asymr - \tilde A_r \qquad , 
\end{equation} 
\begin{equation} 
\label{BC} 
A_{-} \asymr \left(\begin{array}{cc}  
 \frac {\omega_{-}} 2 &  \frac \sigma {2  r} \\ 
  0         &  - \frac {\omega_{-}} 2  
  \end{array}\right) \equiv K_- \, , \quad 
\tilde A_{+} \asymr  \left(\begin{array}{cc}  
 \frac {\omega_{+}} 2 &  0 \\ 
  - \frac \sigma {2 r} &  - \frac {\omega_{+}} 2  
  \end{array}\right) \equiv \tilde K_+ \, , 
\end{equation} 
\begin{equation} 
\label{BCBC} 
A_{+} \asymr \left(\begin{array}{cc}  
 \frac {\omega_{+}} 2  &  \frac  {\stackrel {(2)}{g}_{++}} { r} \\ 
  \frac { r} {\ell^2} + \frac {\sigma_{-+}-\sigma_{+-}}{ r}  
   &  - \frac {\omega_{+}} 2 
  \end{array}\right) \quad , \quad 
\tilde A_{-} \asymr \left(\begin{array}{cc}  
 \frac {\omega_{-}} 2 &  
 \frac {- r} {\ell^2} + \frac {\sigma_{-+}-\sigma_{+-}}{ r}  \\ 
    \frac  {- \stackrel {(2)}{g}_{--}} { r}  &  -  \frac {\omega_{-}} 2 
  \end{array}\right)  \, , 
\end{equation} 
where we have introduced the null components of the connection 2-form 
${\underline \omega}^{12} \equiv \underline \omega =  
\omega_{+} {\underline \theta} ^{+} + 
\omega_{-} {\underline \theta} ^{-}$ and for further convenience 
the notation $K_-$ and $\tilde K_+$. 
Note that the combination 
$(\sigma_{-+}-\sigma_{+-})/{ r}$ can always be canceled by an infinitesimal 
(in the limit $ r \rightarrow \infty$) Lorentz transformation acting 
on $\theta^1$ and $\theta^2$; this gauge freedom explains that this term 
never appears in the subsequent calculations.  
Using these relations, we see that in the large $\bar r$ limit,  
the EH action (\ref{SEH}) 
presents two types of divergences, one in $\bar r^2$ and 
if $\sigma$, i.e.  $\stackrel {(0)} {\cal R}$ (see eq. (\ref{sigR})), 
does not vanish,  
another in $\log{\bar r}$. In contrast, the CS action ($\ref{SCS}$)  
only presents a  $\log{\bar r}$ divergence. The quadratic divergence of  
$S_{EH}$ 
is found in 
${\cal B}_{CS}^{(r)}$. Indeed: 
\begin{equation} 
\label{rbound} 
{\cal B}_{CS}^{(r)} = - \frac {\bar r^2} {8  \, \pi \, G \, \ell^3} 
\int_{|r|=\bar r} \sqrt{\stackrel {(0)} {\bf g}}  \, d\varphi \, dt 
 \, + \, O(\bar r^{\, -2})  
\qquad . 
\end{equation} 
Note that in the limit $\bar r \rightarrow \infty$, this term  
does not contain dynamical degrees of freedom.  
The logarithmic divergence of $S_{EH}$ appears in both $S_{CS}$ and 
${\cal B}_{CS}^{(t)}$, which can be written as: 
\begin{equation} 
\label{BCS2} 
{\cal B}_{CS}^{(t)} = 
\asym   
 \frac \ell {32 \pi G} 
\log \bar r 
\int_{|r|=\bar r}  \stackrel {(0)} {\cal R}  
 \sqrt{\stackrel {(0)} {\bf g}}  \, d\varphi \, dt \qquad . 
\end{equation} 
This term will no longer be considered in this section but discussed  
in section {\bf 6}, where the origin of the anomaly will be analyzed. 
 
As a remark {\it en passant}, let us note that the boundary contribution  
${\cal B}_{CS}^{(r)}$ (\ref{rbound})  
can here be expressed in terms of the  
"extrinsic curvature 2-form'' ${\cal \underline {\underline {K}}}$, 
defined as the product of  
the trace of the extrinsic curvature tensor with the induced  
surface element 2-form. Indeed, on a surface of constant $r$,  
this 2-form is  
given by: 
\begin{equation} 
{\cal \underline {\underline {K}}}= \frac 1 {2 \ell} 
\epsilon_{\mu\nu\rho} \underline \Theta ^\mu \wedge  
\underline \Omega^{\nu\rho} \qquad , 
\end{equation} 
which implies that on this surface: 
\begin{equation} 
\ell \, Tr ({\bf A}\wedge {\bf \tilde A})=  
{\cal \underline{\underline{K}}} \qquad . 
\end{equation} 
Hence, we may introduce a gravity action $S_G$ suitable for Feynman 
path integral, which  
is finite when the boundary metric is flat (and has well-defined 
functional derivatives) \cite{MB}: 
\begin{equation} 
S_G = \frac 1 {16 \pi G} \int_{\cal M} \,  
   ( \, { \cal R} - 2 \, \Lambda ) 
       {\boldsymbol{\eta}} \, + \, \frac 1 {16 \pi G} 
\int_{|r|=\bar r} \, {\cal \underline {\underline {K}}} \quad , 
\end{equation} 
This gravity action differs from that obtained using the 
Gibbons-Hawking procedure \cite{GH}, as  
the term involving the extrinsic 
curvature is equal to half of the usual one \cite{BAN}. This is not 
in contradiction with \cite{GH}, as we are working in the Palatini 
formalism. 
 
Let us now consider the on-shell 
variation of the actions $S_{EH}$ and $S_{CS}$. 
We therefore re-write $S[\bf A]$ (\ref{SA}) as: 
\begin{equation} 
\label{21} 
S[{\bf A}] = \frac {\ell}  {16 \pi G}  
\int_{\cal V}{ Tr (2\ A_t F_{r\varphi} + A 
_{\varphi}\dot A_r -A_r \dot A_{\varphi} )} 
 \, dr \, d\varphi \, dt \, + \, {\cal B}^{(r)}[{\bf A}]  
\qquad,  
\end{equation} 
where the boundary term is: 
\begin{equation} 
{\cal B}^{(r)}[{\bf A}] = - \frac {\ell}  {16 \pi G}  
\int_{|r|=\bar r} Tr (A_t A_\varphi) \, d\varphi\, dt 
\qquad .
\end{equation} 
Using the asymptotic behaviour of the fields ${\bf A}$ and ${\bf \tilde A}$ 
given by eqs (\ref{BC0}, \ref{BC}, \ref{BCBC}), 
we find that at spatial infinity 
($|r|=\bar r \rightarrow \infty$):  
\begin{eqnarray} 
\label{dBC1} 
\delta A_{-} \, &= \, O(r^{-2}) \, &= \, \delta \tilde A_{+} \qquad, \\ 
\label{dBC2} 
Tr (A_{-} \delta A_{+}) \, &= \, O(r^{-2}) \, &= \,  
Tr (\tilde A_{+} \delta \tilde A_{-}) 
 \qquad ,\\ 
\label{dBC3} 
Tr ( \tilde A_{-} \delta A_{+}) \, &= \, O(r^{-2}) \, &= \,  
Tr ( A_{+} \delta \tilde A_{-} ) \qquad . 
\end{eqnarray} 
Accordingly, the variations of both actions 
$S_{EH}$ and $S_{CS}$ vanish:
\begin{equation} 
\delta S_{CS}=
\int_{|r|=\bar r} O(r^{-2}) \, d\varphi\, dt \qquad , \qquad 
\delta S_{EH}=
\int_{|r|=\bar r} O(r^{-2}) \, d\varphi\, dt \qquad . 
\end{equation} 
It is noteworthy that, owing to the boundary conditions  
(\ref{BC}, \ref{BCBC}),  
the variation of the action $S_{CS}$ vanishes by itself,  
without needing to add any extra boundary term \cite{FE}, contrary to
what was sometimes stated.

\subsection{Second step: Wess-Zumino-Witten action}

Let us now return to the action $S_{CS}$ given by eqs (\ref{SCS},\ref{21}). 
The time components $A_t$ and $\tilde A_t $ play the r\^ole of  
Lagrange multipliers and can be 
eliminated from the bulk action by solving the constraint equations  
$F_{r\varphi} = 0$ and $\tilde F_{r\varphi} = 0$. On the cylinder, 
their general 
solutions \cite{EMSS}
are given by gauge transforming non trivial flat connections 
$h_i$ and $\tilde h_i$: 
\begin{equation} 
\label{DEFHO1} 
 A_i =  G^{-1}_1 h_i  G_1 + G^{-1}_1 \partial_i G_1   \qquad , \qquad
\tilde A_i =  G^{-1}_2 \tilde h_i  G_2 + G^{-1}_2 \partial_i G_2   \qquad ,  
\end{equation} 
with $i$ labelling the coordinates $(r, \varphi)$. 
The flat connection components can be chosen as 
$h_r=0$, $\tilde h_r=0$, $h_\varphi= h(t)$ and 
$\tilde h_\varphi= \tilde h(t)$, where $h(t)$ and  
$\tilde h(t)$ 
are $sl(2,\real)$ generators that only depend $t$. The $SL(2,\real)/Z_2$ 
matrices $G_1$ 
and $G_2$ are assumed univoquely defined on the cylinder:  
$G_1(r,\varphi,t)= G_1(r,\varphi+2\pi,t)$  and similarly for $G_2$.  
Instead of using this represenation of $A$ and $\tilde A$, we choose 
to express them in terms of non-periodic group elements  
$Q_1$ and $Q_2$: 
\begin{equation} 
\label{AQ1} 
A_i = Q^{-1}_1 \partial_i Q_1 \qquad , \qquad   
\tilde A_i = Q^{-1}_2 \partial_{i} Q_2 \qquad , 
\end{equation} 
where 
\begin{equation} 
\label{G1} 
Q_1(r,\varphi,t)=\exp[\varphi \, h(t)]\, G_1(r,\varphi,t) 
\quad , \quad  
Q_2(r,\varphi,t)=\exp[\varphi \, \tilde h(t)]\, 
 G_2(r,\varphi,t) 
\quad . 
\end{equation} 
These two representations of the gauge fields are obviously equivalent. 
We choose to use the second one, as it will naturally lead to a 
single  non-periodic Liouville field on 
the spatial boundary.
 
Due to the asymptotic behaviour of $A_r$ and $\tilde A_r$ (\ref{BC0}),  
$Q_1$ and $Q_2$ asymptotically factorize into: 
\begin{equation} 
\label{Q12asym} 
Q_1(\bar r,\varphi,t)\asym 
q_1(\varphi,t) S(\bar r) \qquad , \qquad 
Q_2(\bar r,\varphi,t)\asym 
q_2(\varphi,t) S(\bar r)^{-1} \qquad , \qquad 
\end{equation} 
with\footnote{Note that if the asymptotic behaviours of 
$\bf A$ and $\bf \tilde A$ are interchanged (as discussed 
before eq. (\ref{BC0})), $S(\bar r)$ must
be replaced by $S(\bar r)^{-1}$ in eq. (\ref{Q12asym}).}:
\begin{equation} \label{ExpreS}
S(\bar r) = \left(\begin{array}{cc}  
 \sqrt{\frac {\bar r} \ell}  &  0 \\ 
  0   &   \sqrt{\frac \ell {\bar r}} 
  \end{array}\right) \qquad . 
\end{equation} 
On the other hand, the components $A_t$ and $\tilde A_t$ in the boundary 
actions may be eliminated in terms of $A_\varphi$, $\tilde A_\varphi$, 
$K_{-}$ and $\tilde K_{+}$ using the boundary conditions (\ref{BC}).
The remaining conditions, given by eq. (\ref{BCBC}), 
restrict the matrix elements of $q_1$ and $q_2$, but are difficult to 
implement at this stage, and we choose not to do so for the  
moment. Dropping these conditions (\ref{BCBC}) implies that we  
must add a boundary  
term to the action $S_{SC}$ 
ensuring the vanishing of its on-shell variation, independently 
of (\ref{BCBC}). This modified action reads as: 
\begin{equation} 
\label{SPRCS} 
S^\prime_{CS}=S_{CS} + {\cal B}_{CS'}^{(r)} - {\cal G}_{CS'} \qquad , 
\end{equation} 
The additional dynamical boundary term ${\cal B}_{CS'}^{(r)}$  
is given by: 
\begin{equation} 
{\cal B}_{CS'}^{(r)} 
=  \frac {\ell} {16 \pi G} \int_{|r|=\bar r}   
Tr (  A_{-} \,  A_{+} + \tilde A_{-} \,  \tilde A_{+}) \,  
\theta \, d\varphi\, dt  \qquad ,
\end{equation} 
where  
$\theta= 2 \sqrt {- \stackrel {(0)} {\bf g} } $.
The ${\cal G}_{CS'}$ term contains no degrees of 
freedom in the limit $\bar r \rightarrow \infty$, 
is finite and reads as: 
\begin{equation} 
\label{GCS} 
{\cal G}_{CS'} = \frac {\ell} {16 \pi G}\,\int_{|r|=\bar r}  
  (\, \omega_{+}\omega_{-}  
 \, + \, 
 \frac 1 {\ell^2} \sigma ) \, \theta \, d\varphi\, dt \qquad . 
\end{equation} 
This term is equal to ${\cal B}_{CS'}^{(r)}$ when all  
boundary conditions are imposed, and ensures that the value of the action $S_{CS}$ 
remains unmodified in the limit 
$\bar r \rightarrow \infty$. 
Here and in the following, such boundary terms without degrees on freedom 
will be qualified as geometrical. 
   
Inserting in $S^\prime _{CS}$ the expression of the gauge fields in terms 
of the $Q_i$ matrices (\ref{Q12asym}), we get a sum of four terms: 
\begin{equation} 
\label{SPCS} 
S^\prime _{CS} =  S_{WZWC} + {\cal B}^{(\varphi)}_{WZWC} + {\cal G}_{WZWC}  
- {\cal G}_{CS'} \qquad . 
\end{equation} 
The first is the chiral WZW action: 
\begin{eqnarray} 
\label{WZWC} 
S_{WZWC} &=& - \Gamma[Q_1] +  \frac {\ell} {16 \pi G} \int_{|r|=\bar r} 
   Tr[ \frac 1 { e_{-}^t} \, q^\prime _1 \, 
 ( q_1^{-1} \partial_{-} q_1 - \, 2\, k_{-} ) ] 
    \, dt \, d\varphi \\ 
  && + \, \Gamma[Q_2] - \frac {\ell} {16 \pi G} \int_{|r|=\bar r} 
   Tr[ \frac 1 { e_{+}^t} \, q^\prime _2 \, 
 ( q_2^{-1} \partial_{+} q_2 - \, 2\, \tilde k_{+} ) ] 
    \, dt \, d\varphi \qquad \nonumber , 
\end{eqnarray} 
where the derivatives $\partial _{+}$ and $\partial _{-}$ are taken along the 
vectors $\vec e_{+}$ and $\vec e_{-}$, $q^\prime=q^{-1} \partial_{\varphi}q$, 
$k_{-}=S(r)\, K_{-}\,S(r)^{-1}$, $\tilde k_{+}=S(r)^{-1}\tilde K_{+}\,S(r)$,  
and the bulk WZW action reads as: 
\begin{equation} 
\Gamma [Q] \, = \, \frac {\ell} { 48 \pi G}  
\int Tr [Q^{-1} dQ \wedge Q^{-1} dQ 
\wedge Q^{-1} dQ] \qquad . 
\end{equation} 
The second term: 
\begin{equation} 
\label{BWZWC} 
{\cal B}^{(\varphi)}_{WZWC} =  \frac {\ell}  {16 \pi G} 
\int_{\cal V} Tr [ \partial_\varphi 
(Q_2^{-1} \partial_r Q_2 \, Q_2^{-1} \partial_t Q_2 
 -Q_1^{-1} \partial_r Q_1 \, Q_1^{-1} \partial_t Q_1)] 
\, dr \, d\varphi\, dt  
\end{equation} 
does not vanish 
in general due to the holonomy encoded in $Q_1$ and $Q_2$. 
The last two terms are geometrical; ${\cal G}_{CS'}$ is given by (\ref{GCS}) 
and ${\cal G}_{WZWC}$ by: 
\begin{equation} 
\label{GWZWC} 
{\cal G}_{WZWC} =  \frac {\ell}  {16 \pi G} 
\int_{|r|=\bar r } 
 [\frac {e_+^t}{e_-^t} \, k_-^2 + \frac {e_-^t}{e_+^t} \, \tilde k_+^2] 
 \, \theta \, d\varphi \, dt \qquad . 
\end{equation} 
 
Defining the new variables  
\begin{equation} 
\label{Q12} 
Q=Q_1^{-1} Q_2 \qquad , \qquad 
q=q_1^{-1} q_2 \qquad ,  
\end{equation} 
one of the fields $q_1$ or $q_2$ can be eliminated from the action $S_{WZWC}$  
using its equation of motion. 
Indeed, the variables $q_1$ or $q_2$ only appear in the  
chiral WZW action in quadratic expressions of their derivatives with 
respect to the angular variable $\varphi$. Their equations of motion lead to: 
\begin{eqnarray} 
\label{q1} 
q_1^\prime &=& \theta \, [ {e_{-}^t} \,  
  ( \partial_{+} q q^{-1}  - q \tilde k_{+} q^{-1} ) + 
    {e^t_{+}} k_{-} +  e^t_+ e^t_- q_1^{-1} n_1(t) q_1 ] \qquad , \\ 
\label{q2} 
q_2^\prime &=& \theta \, [{e_{+}^t} \, 
   ( q^{-1} \partial_{-} q   + q^{-1} k_{-} q ) - 
    {e^t_{-}} \tilde k_{+} +  e^t_+ e^t_- q_2^{-1} n_2(t) q_2 ]  \qquad , 
\end{eqnarray} 
where $n_1(t)$ and $n_2(t)$ are $sl(2,\real)$ generators depending only 
on $t$, appearing upon $\varphi$ integrations. It is easy to see that these 
degrees of freedom decouple from $q$ in the action and, using their 
own equation of motion, may be directly set equal to zero. 
The resulting action becomes: 
\begin{equation} 
\label{WZWCWZW} 
S_{WZWC} = S_{WZW} + {\cal B}^{(\varphi,t)}_{WZW} - {\cal G}_{WZWC}  
\qquad , 
\end{equation} 
with the geometrical contribution canceling 
that occurring in eq. (\ref{SPCS}). 
The (non-chiral) WZW action is given by: 
\begin{eqnarray} 
S_{WZW} &=& \Gamma[Q]  
   - \frac {\ell} {8 \pi G} \int_{|r|=\bar r} 
 Tr[ \frac 1 2 q^{-1}\partial_{+} q \, q^{-1}\partial_{-} q  
    +  \partial_{+} q  q^{-1} k_{-}  \nonumber\\ 
\label{WZW} 
   && \qquad \qquad \qquad \qquad   -   
   q^{-1} \partial_{-} q  \tilde k_{+} 
  - \tilde k_{+} q^{-1} k_{-} q ]  
  \, \theta \, dt \, d\varphi \qquad .  
\end{eqnarray} 
The boundary term ${\cal B}^{(\varphi,t)}_{WZW}$ reads as:  
\begin{equation} 
\label{BWZW} 
{\cal B}^{(\varphi,t)}_{WZW} =  \frac {\ell}  {8 \pi G} 
\int_{\cal V} Tr [ \partial_\varphi 
(Q_2^{-1} \partial_{[r} Q_2  Q^{-1} \partial_{t]} Q) + 
\partial_t 
(Q_2^{-1} \partial_{[\varphi} Q_2 Q^{-1} \partial_{r]} Q)] 
 dr\, d\varphi\, dt \quad .  
\end{equation} 

\subsection{Boundary equations of motion as consistency equations}

Let us for a moment focus on the equations of motion (\ref{q1}, \ref{q2}) 
of $q_1$ and $q_2$ as functions of $q$. 
Using the remaining boundary conditions (\ref{BCBC}) and 
the Gauss decomposition for $SL(2,\real)/Z_2$ elements: 
\begin{equation} 
\label{Gaussq} 
q = \left(\begin{array}{cc}
e^{\phi /2} + xy e^{-\phi/2} &x e^{-\phi/2}\\
y e^{-\phi/2} &e^{-\phi/2}
\end{array}\right) \qquad ,
\end{equation}
eqs (\ref{q1}, \ref{q2}) lead to 6 equations. Four of them:  
\begin{eqnarray}
\label{XY}
x=\frac \ell 2 \partial _ {+} \phi & \quad, \quad & 
y=\frac \ell 2 \partial _ {-} \phi \qquad,\\
\label{DXY}
\ell (\partial_{-} + \omega _{-}) x + \frac 1 2 \sigma + e^\phi=0 & \quad, 
\quad & 
\ell (\partial_{+} - \omega _{+}) y + \frac 1 2 \sigma + e^\phi =0 \quad,
\end{eqnarray}
determine $x$ and $y$ as functions of $\phi$ and combine to give:
\begin{equation}
\label{43}
\Box \phi + \frac 8 {\ell^2} e^{\phi} + \frac 4 {\ell^2} \sigma = 0 \qquad .
\end{equation}
This is the Liouville equation on a curved background, the curvature
being given by eq. (\ref{sigR}).
The last two equations are:
\begin{eqnarray}
\label{eq5}
(\partial_{+} + \omega_{+})\partial_{+} \phi - 
 \frac 1 2 ( \partial_{+}\phi)^2 + \frac 2 {\ell^2} \, 
\stackrel {(2)} {g}_{++} & =&  0 
\qquad ,\\
\label{eq6}
(\partial_{-} - \omega_{-})\partial_{-} \phi - 
 \frac 1 2 ( \partial_{-}\phi)^2 + \frac 2 {\ell^2} \, 
\stackrel {(2)} {g}_{--} & =&  0 
\qquad .
\end{eqnarray}
Using eq. (\ref{gasymp}) and the expression of the energy-momentum tensor 
of the Liouville field,
these equations and eq.(\ref{sigR}) can be summarized as:
\begin{equation}
\label{Tab2}
\stackrel {(2)} g _{ab}  =   \frac {\ell^2} 2 \, ( \, T_{ab} 
- \eta _{ab}  \stackrel {(0)}{\cal R} \, ) \qquad .
\end{equation}

\subsection{Third step: Liouville action}

We now return to the non-chiral WZW action (\ref{WZW}). 
Using for the matrix $Q$ a Gauss decomposition in terms of 
$\Phi$, $X$ and $Y$ similar to (\ref{Gaussq}),
the bulk term $\Gamma (Q)$ can be written as a sum of three
kinds of boundary contributions:
\begin{equation}
\Gamma(Q)=- \frac {\ell} {16 \pi G} \int d \wedge [e^{-\Phi} dX \wedge dY]
={\cal B}^{(r)}_\Gamma +  {\cal B}^{(\varphi,t)}_\Gamma
\qquad .
\end{equation}
The contribution on surfaces of 
constant $|r|=\bar r$ yields in the limit $\bar r \rightarrow \infty$:
\begin{equation}
\label{BG}
{\cal B}^{(r)}_\Gamma= \frac {\ell} {8 \pi G} \int_{|r|=\bar r} 
e^{-\phi}( \partial_{[\varphi} y \partial_{t]} x ) \, d\varphi \, dt
 \qquad ,
\end{equation}
where we have used eqs (\ref{Q12}, \ref{Q12asym}), which give
the asymptotic expressions:
\begin{equation}
\label{AXYPHI}
e^{-\Phi}\sim \frac {r^2}{\ell^2} e^{-\phi} \qquad , \qquad
X \sim \frac \ell r x  \qquad , \qquad Y \sim \frac \ell r y 
\qquad .
\end{equation}
The last contribution is:
\begin{equation}
\label{BL}
{\cal B}^{(\varphi,t)}_\Gamma  = \frac {\ell} {8 \pi G} 
\int_{\cal V} [ \partial_\varphi
( e^{-\Phi} \partial_{[t} Y \partial_{r]} X) +
\partial_t
( e^{-\Phi} \partial_{[r} Y \partial_{\varphi]} X)]
 dr\, d\varphi\, dt \qquad .
\end{equation}
Accordingly, the non-chiral WZW action (\ref{WZW}) may be written as:
\begin{equation}
\label{WZWXY}
S_{WZW}= S_{(x,y,\phi)} + {\cal B}^{(\varphi,t)}_\Gamma 
+ {\cal G}_{(x,y,\phi)}\qquad ,
\end{equation}
where 
\begin{eqnarray}
\label{SXYPHI}
S_{(x,y,\phi)}&=& - \frac {\ell}  {16 \pi G}
 \int_{|r|=\bar r }[ \frac 1 2 \partial_+ \phi \partial_- \phi 
+ \omega_-  \partial_+ \phi - \omega_+  \partial_- \phi \qquad \qquad
\nonumber \\
&& \quad \quad 
 + 2  e^{-\phi} (\partial_- x +\omega_- x +\frac {\sigma}{2 \ell})
(\partial_+ y - \omega_+ y +\frac {\sigma}{2 \ell})] \theta \, d\varphi \, dt
\, , 
\end{eqnarray}
and the geometrical term ${\cal G}_{(x,y,\phi)}$ is:
\begin{equation}
\label{GXY}
{\cal G}_{(x,y,\phi)}=  \frac {\ell}  {16 \pi G} \int_{|r|=\bar r }
\omega_+ \omega_- \theta \, d\varphi \, dt \qquad .
\end{equation}
This term cancels the first term in the geometric contribution
appearing upon modifying the CS action, see eqs (\ref{SPRCS},\ref{GCS}). 
 
Finally, the action $S_{(x,y,\phi)}$ (\ref{SXYPHI}) can be expressed 
in terms of $\phi$ only by eliminating \cite{HT,CHD} the variables
$x$ and $y$ in terms of the constants
of motion defined by (\ref{DXY}), using the same trick as the one that leads
to the Maupertuis action in classical mechanics with conserved energy.
From the definition of the connection and eqs (\ref{XY}, \ref{DXY}), we find:
\begin{eqnarray}
\theta ( \partial_- x + \omega_- x ) 
&=& \partial_\varphi (\theta ^+_t x) - \partial_t (\theta ^+_\varphi x)
= - \theta ( \frac {e^\phi} \ell + \frac \sigma {2\, \ell})
\qquad ,\nonumber \\
\theta ( \partial_+ y - \omega_+ y ) 
&=& - \partial_\varphi (\theta ^-_t y) + \partial_t (\theta ^-_\varphi y)
= - \theta ( \frac {e^\phi} \ell + \frac \sigma {2\, \ell} )
\qquad .
\end{eqnarray}
Using these equations, we add to the action $S_{(x,y,\phi)}$ zero written as:
\begin{eqnarray}
0 &=&  \frac {\ell}  {8 \pi G}
 \int_{|r|=\bar r } 
\left[e^{-\phi} (\partial_-x + \omega_- x)
           (\partial_+y - \omega_+y +\frac \sigma {2 \ell})\right .
\nonumber \\
&& \left . \qquad \qquad 
+ e^{-\phi} (\partial_+y - \omega_+ y)
          (\partial_-x + \omega_-x +\frac \sigma {2 \ell})\right ] 
\theta d\varphi dt
\nonumber \\ 
&& + \frac {\ell}  {8 \pi G}
 \int_{|r|=\bar r } 
 \left [\partial_\varphi (\theta^+_t x -\theta^-_t y)
  - \partial_t (\theta^+_\varphi x -\theta^-_\varphi y) \right] d\varphi dt
\qquad .
\end{eqnarray}
Adding this null expression to  $S_{(x,y,\phi)}$ allows to eliminate
$x$ and $y$ as functions of $\phi$ using eqs (\ref{XY}, \ref{DXY}),
while keeping the correct $\phi$ field equations and holonomy
contributions. Accordingly, we get:
\begin{equation}
\label{XYL}
S_{(x,y,\phi)} = S_L + {B}_L + {\cal G}_L \qquad ,
\end{equation}
where $S_L$ is the Liouville action on curved background:
\begin{equation}
\label{SL}
S_L= - \frac {\ell} {32 \pi G} \int_{|r|=\bar r} 
[\frac 1 2 \stackrel {(0)} g 
\strut ^{ab}
\partial_a \phi \partial_b \phi - \frac 8 {\ell^2} e^{\phi}+ 
\stackrel {(0)}{\cal R} \phi] 
 \, \sqrt{- \stackrel {(0)} {\bf g}} \, dt \, d\varphi \qquad .
\end{equation}
Let us emphasize that the curvature term
appearing here comes directly from its definition in terms of the asymptotic
metric $\stackrel {(0)} {\bf g}$, and not through $\sigma$ as it is the case 
in eq. (\ref {43}). 
The term  ${ B}_L$ is defined on the ${r, \varphi}$ and  
${r, t}$ boundaries:
\begin{equation}
\label{BBL}
{B}_L= \frac {\ell} {16 \pi G} \int _{|r|=\bar r}
\left [
\partial_a (\sqrt{- \stackrel {(0)} {\bf g}} \stackrel {(0)} g \strut ^{ab}
\partial_b \phi) 
+ \partial_\varphi (\omega_t \phi) -
 \partial_t (\omega_\varphi \phi) \right ] d\varphi dt \qquad ,
\end{equation}
and the geometrical term is given by:
\begin{equation}
\label{GL}
{\cal G}_L = \frac \ell {8 \pi G} \int _{|r|=\bar r} \frac \sigma {\ell^2}
\theta d\varphi dt \qquad .
\end{equation}

\section{Holonomies}

In the previous section, we have shown that the EH action can be
expressed as a sum of terms defined on the $r$, $\varphi$ and $t$
boundaries without any remaining bulk terms. The dynamical equation
resulting from imposing the stationarity of the action on the spatial
boundary $|r|=\bar r$, is that of Liouville on a curved background.
We now analyze the dynamical content of the $\varphi$-boundary terms
${\cal B}^{(\varphi)}$ that occur when $\Sigma $ has the topology of the 
cylinder.

The first two non vanishing holonomy terms are given by
eqs (\ref{BWZWC}, \ref{BWZW}) and appear when going from
the CS action to the non-chiral WZW action. Together they give, 
after substitution of $Q_1$ by $Q_2Q^{-1}$:
\begin{equation}
\label{BWZW1}
{\cal B}^{(\varphi)}_{WZWC} + {\cal B}^{(\varphi)}_{WZW}=
\frac \ell {16 \pi G} \int_{\cal V} \partial_\varphi
\left [ 2 \, Q_2^{-1} \partial_r Q_2 Q^{-1} \partial_t Q -
         Q^{-1} \partial_r Q Q^{-1} \partial_t Q \right ] 
\, dr \, d\varphi \, dt \, .
\end{equation}
Contrary to what happens on the boundary $|r|=\bar r$,
$Q_2$ cannot be eliminated in terms of $Q$ by using its equation of motion.
When going from the non-chiral WZW action to the $S_{(x, y, \phi)}$
action, another holonomy term
appears, ${\cal B}^{(\varphi)}_{\Gamma}$,
given by eq. (\ref{BG}). Adding this term to (\ref{BWZW1}),
we obtain the $\varphi$-boundary action ${\cal B}^{(\varphi)}$:
\begin{equation}
\label{BPHI}
{\cal B}^{(\varphi)}= \frac \ell {16 \pi G} \int_{\cal V} \partial_\varphi
\left [ 2 \, Q_2^{-1} \partial_r Q_2 Q^{-1} \partial_t Q 
 - \frac 1 2 \partial_t \Phi \partial_r \Phi - 2  \partial_t X \partial_r Y
 e^{-\Phi} \right ] \, dr \, d\varphi \, dt \qquad , \qquad
\end{equation}
which encodes all bulk terms that survive in addition to the 
Liouville action  (\ref{SL}) on the spatial $|r|=\bar r$ boundary.

The equations of motion on the $\varphi$-boundary can be obtained from this
action by varying $Q_2$, $X$, $Y$ and $\Phi$. But it is easier to obtain 
them by varying $Q_1$ and $Q_2$ in the CS action (\ref{SPCS}) 
expressed in terms of these fields. We get:
\begin{equation}
\delta \, S'_{CS}= 
\frac \ell {8 \pi G} \int_{\cal V} \partial _\varphi
Tr[Q_1^{-1} \partial_r(\partial_t Q_1 Q_1^{-1}) \delta Q_1 -
 Q_2^{-1} \partial_r(\partial_t Q_2 Q_2^{-1}) \delta Q_2] 
dr\,d\varphi\,dt \qquad. 
\end{equation}
In terms of the periodic group elements $G_1$ and $G_2$ defined in eqs
(\ref{G1}), which factorize in the same way as the non-periodic ones
(cf eq. \ref{Q12asym}):
\begin{equation}
G_1(\bar r,\varphi,t)\asym
g_1(\varphi,t) S(\bar r) \qquad , \qquad
G_2(\bar r,\varphi,t)\asym
g_2(\varphi,t) S(\bar r)^{-1} \qquad , \qquad \label{G12asym}
\end{equation}
the variation of the action  becomes:
\begin{equation}
\delta \, S'_{CS}= 
\frac \ell {8 \pi G} \int_{|r|=\bar r} 
Tr[\partial_t g_1 \, g_1^{-1}\, \delta \zeta -
\partial_t g_2 \, g_2^{-1} \,  \delta \tilde \zeta  \, ] |_{\varphi=0} \, 
dt \qquad ,
\end{equation}
with $\delta \zeta(t)=\exp(-2 \pi h)\delta \exp(2 \pi h)$ and similarly
for $\delta \tilde \zeta$.
The equations of motion are thus:
\begin{equation}
\label{holog}
\partial_t g_1 \, g_1^{-1} \,|_{\varphi=0} |_{r=-\bar r}^{r=\bar r}=0 \qquad , \qquad
\partial_t g_2 \,g_2^{-1} \, |_{\varphi=0} |_{r=-\bar r}^{r=\bar r}=0 \qquad ,
 \qquad
\end{equation}
whose solutions are simply
\begin{eqnarray}
\label{MOTHOL}
g_1( \varphi=0,\, t)|_{r=-\bar r}&=&
g_1( \varphi=0,\, t)|_{r=+\bar r} \,\, {\cal C}_1
\qquad,\qquad \nonumber\\
 g_2( \varphi=0,\, t)|_{r=-\bar r}&=& 
g_2(\varphi=0,\, t)|_{r=+\bar  r} \,\, {\cal C}_2
\qquad,
\end{eqnarray}
with ${\cal C}_1$ and ${\cal C}_2$ constant matrices.
Hence, the bulk field configuration does not appear in the equations of motion;
only the fields defined on the spatial boundary do. 

It is interesting at this stage to compare these results with the equivalent 
ones obtained in \cite{HMS}
on the basis of expression (\ref{DEFHO1}) instead of
(\ref{AQ1}) for the gauge potentials. In the latter approach, 
the CS action reduces to a sum of actions on the spatial boundaries
involving
globally defined (single-valued) fields which are coupled to the flat 
connections $h$ and $\tilde h$ and hence to each other.
In contrast, our analysis based on multi-valued fields appears to be simpler,
as the only remaining coupling between the fields in the equations
of motion is through
eqs (\ref{MOTHOL}).

\section{Flat boundary metric}

To further analyze the significance of the holonomy contributions taken 
into account in the previous section via $\varphi$-boundary terms, we compute 
the gravitational holonomy encoded in the solutions of Einstein's equations. 
For this purpose, we restrict 
our analysis to the flat $\stackrel {(0)} {\bf g}$ metric (infinite cylinders
have no moduli). For such asymptotic geometry, 
the expansion described in
eq. (\ref{FGTH}) stops at order $r^{-1}$
and the complete expression of locally AdS metrics reads as \cite{BAN}:
\begin{equation}
\label{flatmet}
ds^2 = \ell^2 \frac {dr^2} {r^2} +\left( \frac r \ell\, d x^{+} + 
  \frac \ell r \, L_-(x^-) \, d x^-\right)\left(
\frac r \ell \, d x^{-} + 
     \frac \ell r \, L_+(x^+) \, d x^+ \right)
\qquad ,
\end{equation}
This metric describes two asymptotic regions: one in the neighbourhood of
$r=\infty$, the other near $r=0$. In order to have a single radial coordinate
that leads to the FG metric expression on both asymptotic
regions simultaneously we have to introduce a new radial 
coordinate $\rho$ defined by:
\begin{equation}
\rho=r\qquad \mbox{\rm for}\ r>r_2 \qquad , \qquad
\rho=-\frac{\ell^2}r \qquad \mbox{\rm for}\ r<r_1 \qquad ,
\end{equation}
where $r_1<r_2$ are two arbitrarily chosen positive values of $r$; between
them, $\rho$ is given by any smoothly interpolating increasing function of
$r$. So, near $\rho=\infty$ we may choose as frame:
\begin{equation}
\Theta^0=\ell\,\frac {d\rho}{\rho}\, ,\quad \Theta^+=\frac \rho\ell \,dx^+ +
\frac\ell\rho \,L_-(x^-)\, dx^-\, ,\quad  \Theta^-=\frac \rho\ell \,dx^- +
\frac\ell\rho \,L_+(x^+)\, dx^+\, .
\end{equation}
It continues near $\rho=-\infty$ into:
\begin{equation}
\Theta^0=-\ell\,\frac {d\rho}{\rho} \, ,\quad 
\Theta^+=-\frac\rho\ell\,L_-(x^-)\, dx^- -\frac \ell\rho \,dx^+  \, ,\quad 
 \Theta^-=-
\frac\rho\ell\,L_+(x^+)\, dx^+ - \frac \ell\rho \,dx^- \,.
\end{equation}
To avoid closed causal curves we must assume the functions
$L_-(x^-)$ and $L_+(x^+)$ to be non-negative. In this case the geometry
(\ref{flatmet}) presents a singularity on the surface $r^4=\ell^4\, L_-(x^-)
L_+(x^+)$. Such a singularity deserves special attention: in the 
case of BTZ black
holes it corresponds to the troath of the Einstein-Rosen bridge, but it is not
clear if it still corresponds to a coordinate singularity in
the more general case or becomes a true singularity.

On the boundary $r=\infty$, the Liouville field solution of 
eq. (\ref{43}) can be expressed as:
\begin{equation}
\phi= \log \left \vert \frac {\ell^2 \,  f^{\prime}_+ \,  f^{\prime}_-}
{(f_++ f_-)^2} \right \vert \qquad ,
\end{equation}
where $f_+$ (resp. $ f_-$) is a function of $x^+$ (resp. $x^-$) only and $
f^{\prime}_+$ (resp. $ f^{\prime}_- $) its derivative with respect to its
argument. 
To yield the metric (\ref{flatmet}) these functions 
must be related to the components of the subdominant term of the asymptotic
metric as:
\begin{equation}
\label{gf}
L_+= \frac {\ell^2} 2 \left [
\frac 3 2 \left ( \frac {f''_+}{ f'_+} \right ) ^2
-  \frac { f'''_+}{ f'_+} \right ] \qquad , \qquad 
L_-= \frac {\ell^2} 2 \left [
\frac 3 2 \left ( \frac {f''_-}{ f'_-} \right ) ^2
-  \frac { f'''_-}{ f'_-} \right ] \qquad .
\end{equation}
Here we see the important role played by the local character of
the gauge fields. Indeed,
though $L_+(x^+)$ and $L_-(x^-)$ are periodic in their arguments,
the functions $f_\pm$, and thus the Liouville field $\phi$, 
are not necessarily so. 

Let us consider the right handed
sector. For a given $L_+$ function,
the solution $f_+$ of eq. (\ref{gf}) can be obtained 
by posing $f'_+ = 1/w_+^2$,
the function $w_+$ being a solution of the linear second order equation:
\begin{equation}
w_+''=\frac{L_+(x^+)} {\ell ^2}\, w_+ \qquad . \label{weq}
\end{equation}
Floquet's theory \cite[sec. {\bf 19.4}]{WW} gives us the general 
functional
form of the solutions of this equation. These 
can be written as linear combinations of 
particular solutions given by the product of 
exponential and periodic functions:
\begin{equation}
\label{EXPER}
{\cal F}(x^+)=\exp{(\mu_+\, x^+)}F(x^+)  \quad {\rm and} \quad 
{\cal K}(x^+)=\exp {(-\mu_+\, x^+)} K(x^+) 
\quad  , 
\end{equation}
where $\mu_+$ is a real or purely imaginary constant. 
As a consequence,
 the function $f_+$ 
appearing in  eq. (\ref{gf}) can, for real values of $\mu_+$,
 always be chosen as the product of
$\exp{(2\mu_+\,x^+)}$ with a periodic function; but of course more
 complicated functional
forms are also possible and are even
required when $\mu_+$ is purely imaginary. 
The most general expression of $f_+$ depends on
three parameters, and is obtained as the integral of the inverse of the square
of the general solution of eq. (\ref{weq}).

The discontinuities of the Liouville field (of the function $f_\pm$) 
when their argument increases by $2\, \pi$
reflect an important geometrical
property of asymptotic $\rm AdS_3$ spaces. They encode in a simple way
the global holonomy properties of the general metric (\ref{flatmet}).
We now attempt to clarify this.

In the $SL(2,\real)$ basis, the equation of parallel
transport by means of the connections $\bf A$ is:
\begin{equation}
\label{eqpartrans}
\partial_\mu \Psi + A_\mu^a J_a \Psi =0 \qquad ,
\end{equation}
where $\Psi$ has the two components $\Psi_1$ and $\Psi_2$.
Note that
these equations, together with those of the left handed sector,
 are exactly those defining the 
Killing spinors of the manifold. The general solutions of eq. 
(\ref{eqpartrans})
for the metric (\ref{flatmet}) are of the form:
\begin{equation}
\label{psi1psi2}
\Psi_1= - \sqrt{\frac \ell r} \ell \,  \psi'_+(x^+) \qquad , \qquad 
\Psi_2= \sqrt{\frac r \ell} \psi_+(x^+) 
\qquad , 
\end{equation}
with the chiral field $\psi_+(x^+)$ satisfying:
\begin{equation}
\label{SPK}
\psi''_+ = \frac {L_+} {\ell^2} \psi_+
\qquad .
\end{equation}
This is nothing else but equation (\ref{weq}) defining $f_+$.
Hence, the general solutions of eq. (\ref{SPK}) can be
written as a linear combination of $1/\sqrt{f'_+}$ and
$f_+/\sqrt{f'_+}$. We find:
\begin{equation}
\left(\begin{array}{c} \Psi_1(r,x^+) \\ \Psi_2(r,x^+) \end{array}\right)
=P(r,x^+;r_0,x^+_0)\left(\begin{array}{c} \Psi_1(r_0,x^+_0) \\ \Psi_2(r_0,x^+_0) \end{array}\right)
\qquad ,
\end{equation}
where the parallel transport matrix $P(r,x^+;r_0,x^+_0)$ is given by:
\begin{equation}
P= S^{-1}(r) \, p \, S(r_0)   \qquad , 
\end{equation}
with (we drop the '+' subscript to lighten the notation):
\begin{equation}
\label{holomat} 
p=
 \left(\begin{array}{lcr} 
\sqrt{\frac { f'}{ f'_0 }}
\left[1 + \frac {f''(f_0-f)}{2 \, (f')^2} \right] & &
\frac{\ell^2}{4} \left[ 
\frac{ 2({f'}_0^2 f'' - {f'}^2 f''_0) + f''_0 f'' (f-f_0)}
{ (f' f'_0)^{3/2}} \right] \\ & & \\
\frac{1}{\ell^2} \frac {(f_0-f)}{\sqrt{f'_0 f'}} & &
\sqrt{\frac { f'_0}{f' }}\left[1+
 \frac {f''_0 (f-f_0)}{2 \,(f_0')^2 }\right]  \end{array}\right) \qquad ,
\end{equation}
whose eigenvalues are simply $\exp(\pm 2 \pi \mu_+)$.
Thus, since $\mu_+$ is given by the zeros of an infinite determinant
built from all the Fourier coefficients of $L_+$ 
\cite[sec. {\bf 19.4}]{WW}, the holonomies
depend on all the on-shell values of the classical generators of the 
asymptotic Virasoro algebra.
The matrix $\tilde p(r,x^-;r_0,x^-_0)$ giving 
the $\tilde \Psi$ solution can be 
obtained from $p^{T-1}$ by the substitutions $f_+ \rightarrow f_-$ and
$x^+\rightarrow x^-$.

The matrix $p$ immediately yields the Wilson loop matrix
$S^{-1}(r_0) \, w \, S(r_0)$
of parallel transport around close loops, by posing 
$x^+=x^+_0+2 \pi \ell$. They also yield the matrices $q_1$ and $q_2$
defining the flat connection. Indeed, inserting
eqs (\ref{AQ1}) into eq. (\ref{eqpartrans}), we obtain:
\begin{equation}
q_1(x^+)=q_1(x^+_0)\,p^{-1} \qquad , \qquad
q_2(x^-)=q_2(x^-_0)\, \tilde p^{-1}  \qquad .
\end{equation}

To make the decomposition (\ref{G1}) explicit,
we use as set of solutions of eq. (\ref{SPK}) the
following linear combination of the quasi-periodic solutions (\ref{EXPER}):
\begin{eqnarray}
U(x^+)&=& \ell\left[{\cal K}'_0 \, {\cal F}(x^+)- {\cal F}'_0 \, {\cal K}(x^+)
\right]  \qquad ,\\
V(x^+)&=&  {\cal F}_0 \, {\cal K}(x^+)- {\cal K}_0 \, {\cal F}(x^+)
\qquad ,
\end{eqnarray}
where the subscript $0$ indicates that the functions must be taken at an
arbitrary reference point $x_0^+$; the functions $\cal F$ and $\cal K$ are 
(partially) normalized so that their wronskian is $1/ \ell$. This implies that
\begin{equation}
U(x^+_0)=1 \quad , \quad U'(x^+_0)=0 \quad , \quad 
V(x^+_0)=0\quad , \quad V'(x^+_0)=\frac 1 \ell \quad .
\end{equation}
and the 
wronskian of $U$ and $V$ is also equal to $1/\ell$:

In terms of these solutions,
we can write explicitly the
non-trivial flat connection $h(t)$ generating the holonomies (eq. \ref{G1}). 
When $\mu_+$ is real we have:
\begin{equation}
h(t)= 2\, {\mu_+}  \, q_1(x^+_0)\, \left(\begin{array}{rr}
\ell{\cal K}'_0&-\ell {\cal F}'_0 \\
-{\cal K}_0&{\cal F}_0
\end{array}\right)J_0\, \left(\begin{array}{rr}
{\cal F}_0&\ell {\cal F}'_0\\
{\cal K}_0& \ell {\cal K}'_0
\end{array}\right)\,q_1^{-1}(x^+_0) \qquad .
\end{equation}
Thus $h(t)$ is a just a similarity transform of the 
matrix $2\, \mu_+ \, J_0$.
As a consequence, the matrix $g_1$ (see eq. \ref{G12asym}), 
which depends on both $t$ and $\varphi$ (and not only on $x^+$),  
is globally defined (as it must be), its entries being given by
products of exponentials of $\pm \mu_+\, t$ and periodic functions of $x^+$.
When $\mu_+$ is purely imaginary, we pose $\mu_+= -i\nu_+$, 
and see that the generators of holonomies $h(t)$ are now
similarity transforms of $2\,\nu_+\, J_2$, thus of
a timelike generator, whereas
the generator found for real $\mu_+$  is spacelike.

Finally note that the expressions (\ref{psi1psi2}) and the
subsequent developments could also be obtained by
limiting ourselves to an asymptotic evaluation of eqs (\ref{eqpartrans}).
This is due to the special form (\ref{flatmet}) of the metric, for which the 
connection form $\underline {\Omega}^{+-}$ vanishes to all orders.
As a consequence, the factorization (\ref{Q12asym}) 
is valid up to $r\mapsto 0$, and eqs (\ref{holog}) are trivially satisfied,
i.e. ${\cal C}_1=1$  and ${\cal C}_2=1$.

To illustrate the above results, let us consider the special case of BTZ black 
holes. From its metric written in FG form (\ref{BTZFG}), 
we immediately obtain the subdominant metric 
components near $r=\infty$ in the null frame:
\begin{equation}
\stackrel {(2)} g_{++}\equiv L_+= \frac 1 4 (M + \frac J \ell)  \qquad , \qquad
\stackrel {(2)} g_{--}\equiv L_-= \frac 1 4 (M - \frac J \ell)  
\qquad .
\end{equation}
For such constant components, eq. (\ref{gf}) is easily
integrated in terms of purely exponential solutions:
\begin{equation}
\label{f1}
f_+ = \frac  {a\, e^{2 \mu _+ \, x^+}+b}
             {c\, e^{2 \mu _+ \, x^+}+d}
\qquad  ,
\end{equation}
where $\mu_+=\frac 1 2 \sqrt{M+ \frac J \ell}$
and $a\,d-b\,c=1$;
$f_- $  is obtained by the substitutions $x^+ \rightarrow x^-$
and $J \rightarrow -J$.
As expected, the Liouville field constructed from these solutions
is in general not globally defined. 
Only for $J=0$ can $\phi$ be extended periodically (it becomes
$\varphi$ independent)  on the interval
$[0, 2 \pi]$ of the $\varphi$-coordinate if we choose $f_+= \exp (\mu \,x^+)$ and
 $f_-= \exp (\mu \,x^-)$, $\mu =\sqrt{M}/2$.

Furthermore, we easily obtain (compared with \cite{Ma} where the same
calculation was originally performed in another context)
the holonomy matrix $w$
corresponding to a Wilson loop surrounding the throat of the 
Einstein-Rosen bridge:
\begin{equation}
w= \left(\begin{array}{cc} 
\cosh \left (  \pi  \sqrt{M+\frac J \ell}\right) &   
- \frac { \sqrt{M+\frac J \ell}}2 
\sinh \left (\pi  \sqrt{M+\frac J \ell}\right) \\
- \frac 2 { \sqrt{M+\frac J \ell}}  
\sinh\left ( \pi  \sqrt{M+\frac J \ell}\right) & 
\cosh \left( \pi  \sqrt{M+\frac J \ell}\right)
 \end{array}\right) \qquad ,     
\end{equation}
and similarly for $\tilde w$ with
$J \rightarrow -J$. The holonomy is thus always non-trivial, and 
there exists no globally well defined Killing spinor fields, except
in two cases. The first is
$J=0$ and $M=-1$, in which case
$w=-2 \, J_0 $ and the geometry is
that of the usual  $\rm AdS_3$ space. The second case is
$J=0$ and $M=0$, where we have to impose $\Psi_1=0=\tilde \Psi_2$ in
order to obtain global Killing spinor fields.
In the former case, we have thus 4 Killing vector fields, and in the latter 2,
in agreement with \cite{CH}.

\section{Anomalies}

As a preliminary to the discussion of the link between the diffeomorphism
anomaly of the gravitational action and the conformal anomaly of the
Liouville action, let us briefly summarize the various steps that led 
in section
{\bf 3} from the EH action to  the Liouville action, 
through the CS action. Using eqs (\ref{SCS}, \ref{BEHCS}), 
and assuming the metric to be globally defined on $\cal V$, we find that
the EH action (\ref{SEH}) and CS action (\ref{SCS}) are related by:
\begin{equation}
\label{SEHSCS}
S_{EH}=S_{CS}+{\cal B}_{CS}^{(r)} +{\cal B}_{CS}^{(t)} \qquad ,
\end{equation}
where ${\cal B}_{CS}^{(r)}$ is an $\bar r^2$-divergent term defined on the 
spatial boundary given by eq. (\ref{rbound}), and ${\cal B}_{CS}^{(t)}$
is a $\log{\bar r}$-divergent term corresponding to a total derivative 
with respect to $t$ and given by eq. (\ref{BCS2}).
The EH action presents an $\bar r^2$-divergence, whereas the CS action 
is at most logarithmically divergent.

Furthermore, using eqs 
(\ref{SPRCS},\ref{SPCS},\ref{WZWCWZW},\ref{WZWXY},\ref{XYL}), 
we find the relation between the CS action (\ref{SCS}) and the
Liouville action (\ref{SL}):
\begin{equation}
\label{SCSSL}
S_{CS}=S_L+ {\cal B}^{(\varphi)} + {\cal B}^{(t)} + B 
\qquad .  
\end{equation}
${\cal B}^{(\varphi)}$ is defined
in eq. (\ref{BPHI}) and corresponds to the sum of all 
$\varphi$-boundary terms encountered when going from the CS action to the
Liouville action. Similarly, ${\cal B}^{(t)}$ is the sum of all
$t$-boundary terms and is given by:
\begin{eqnarray}
{\cal B}^{(t)} &=& {\cal B}^{(t)}_{WZW}+{\cal B}^{(t)}_\Gamma 
 \nonumber \\ 
\label{BT}
&=&
\frac {\ell}  {8 \pi G}
\int_{\cal V} \partial_t
\left [ Tr (Q_2^{-1} \partial_{[\varphi} Q_2 Q^{-1} \partial_{r]} Q )
+ e^{-\Phi} \partial_{[r} Y \partial_{\varphi]} X \right ]
 dr\, d\varphi\, dt 
\qquad ,
\end{eqnarray}
with ${\cal B}^{(t)}_{WZW}$ and ${\cal B}^{(t)}_\Gamma$ defined by eqs 
(\ref{BWZW}, \ref{BL}).
Finally, the term $B$ is sum of all geometrical terms
(\ref{GCS}, \ref{GXY}, \ref{GL}) and the boundary term $B_L$ (\ref{BBL}). 
It is equal to:
\begin{eqnarray}
B &=& 
\frac {\ell} {16 \pi G} \int _{|r|=\bar r}
\partial_\varphi \left [
\sqrt{- \stackrel {(0)} {\bf g}} \stackrel {(0)} g \strut ^{\varphi b}
\partial_b \phi) +
 \omega_t \phi \right ] \nonumber \\
\label{Bdroit}
&& \qquad \qquad 
+ \partial_t \left [
\sqrt{- \stackrel {(0)} {\bf g}} \stackrel {(0)} g \strut ^{tb}
\partial_b \phi) 
- \omega_\varphi \phi  + 2\, \omega_\varphi \right] d\varphi dt \qquad .
\end{eqnarray}
This term is defined on the $(r, \varphi)$ and $(r, t)$ boundaries.

We now analyze the variation under Weyl transformations and under
diffeomorphisms of the EH, CS and Liouville actions. 
Consider a diffeomorphism generated by the vector field
$\bf \xi$, which moves the boundary $|r|=\bar r$
and keeps the metric in 
the FG form (\ref{FG}). Its radial component is asymptotically of the form
${\xi^r}= \delta \alpha(\varphi, t) \, r$, where we assume
$\delta \alpha$
and all its derivatives vanishing on the space-like  boundaries of 
$\cal V$; the $t$ and $\varphi$ components may be assumed to be of order $r^{-2}$\cite{ISTY,BERS}.  
Under the action of this  
diffeomorphism, the metric varies as:
\begin{eqnarray}
\label{difG}
\delta _{D}\stackrel {(0)} g_{ab} &=&  2 \stackrel {(0)} g_{ab} 
\delta \alpha \qquad , \\
\delta _{D}\stackrel {(2)} g_{ab} &=&  \ell^2
\delta \alpha_{,b;a} \qquad .
\end{eqnarray}
The on-shell variation of the EH action (\ref{SEH})
under these transformations can be computed as:
\begin{equation}
\delta_{D} S_{EH} =
\frac  \ell {16 \pi G} 
\int_{|r|=\bar r} Tr [ \xi^r \{ 6 \,\partial_{[r} (\tilde A_\varphi A_{t]}) - 
 2 \, A_r (A_{[\varphi}  A_{t]} + \tilde A_{[\varphi} \tilde A_{t]})\} ]
\, d\varphi dt
\quad .
\end{equation}
Inserting the boundary conditions (\ref{BC0}, \ref{BC}, \ref{BCBC}) in
this equation, we obtain by expanding the metric:
\begin{equation}
\label{dDSEH}
\delta_{D} S_{EH} =
 \frac \ell {16 \pi G}
\int_{|r|=\bar r} [\stackrel {(0)} {\cal R}- \frac {4 \bar  r^2}{\ell^4}]
 \sqrt{\stackrel {(0)} {\bf g}} \, \delta \alpha
 \, d\varphi dt
\qquad ,
\end{equation}
in agreement with \cite{HS, B, BERS}. Similarly, 
the on-shell variation of the CS action (\ref{SCS}) is given by:
\begin{equation}
\delta_{D} S_{CS} =
- \frac  \ell {16 \pi G} 
\int_{|r|=\bar r} Tr [2\,  \xi^r \, A_r (A_{[\varphi}  A_{t]} + 
\tilde A_{[\varphi} \tilde A_{t]}) ]
\sqrt{\stackrel {(0)} {\bf g}}\,
 \, d\varphi dt \qquad ,
\end{equation}
and using the boundary conditions (\ref{BC0}, \ref{BC}, \ref{BCBC}):
\begin{equation}
\delta_{D} S_{CS} = 
 \frac \ell {32 \pi G}
\int_{|r|=\bar r} \stackrel {(0)} {\cal R}
 \sqrt{\stackrel {(0)} {\bf g}} \, \delta \alpha
 \, d\varphi dt
\qquad .
\end{equation}
Using (\ref{SCS}), we see that the difference between the variations 
of the EH and CS actions is transferred in
the ${\cal B}_{CS}$ term (\ref{BEHCS}). We find indeed:
\begin{eqnarray}
\delta_{D} {\cal B}_{CS} &=&
\frac  \ell {16 \pi G} 
\int_{|r|=\bar r} Tr [ \xi^r \{ 6 \,\partial_{[r} (\tilde A_\varphi A_{t]})]
 \, d\varphi dt \\
&=&
 \frac \ell {16 \pi G}
\int_{|r|=\bar r} [\frac 1 2 \stackrel {(0)} {\cal R}
- \frac {4 \bar  r^2}{\ell^4}]
 \sqrt{\stackrel {(0)} {\bf g}} \, \delta \alpha
 \, d\varphi dt
\qquad ,\quad .
\end{eqnarray}
Thus, the variations of the EH and SC actions under diffeomorphisms  
are related by:
\begin{equation}
\delta_{D} S_{EH} =  2 \, \delta_{D} S_{CS}
+ \frac {\bar  r^2} {4 \pi G \ell^3}  
\int_{|r|=\bar r} \sqrt{\stackrel {(0)} {\bf g}} \delta \alpha
 \, d\varphi dt \qquad .
\end{equation}
The variation $\delta_{D} S_{EH}$ presents a $\bar r^2$ divergence whereas 
$\delta_{D} S_{SC}$ is finite. More strikingly, the finite contributions
of $\delta_{D} S_{EH}$ and $\delta_{D} S_{SC}$
differ by a factor of 2. 

Since the Liouville action $S_L$ (\ref{SL}) is finite and defined on the
spatial boundary $|r|=\bar r$, it is invariant 
under the 3-dimensional diffeomorphisms. 
Hence, when going from the CS action to the Liouville action, 
the variation under diffeomorphisms must be totally transferred
to the additional terms given in eq. (\ref{SCSSL}), in particular
to the $\cal B$ boundary terms, 
as the $B$ term (\ref{Bdroit}) is invariant under 3-dimensional 
diffeomorphisms.
It is easy to check that it is transferred to 
the ${\cal B}^{(t)}$ term (\ref{BT}) that presents a logarithmic divergence 
in $\bar r$.
We find indeed that on-shell:
\begin{equation}
{\cal B}^{(t)} \asym
 - \frac \ell {16 \pi G} \int_{\cal V} \frac 1 r \partial_t \omega_\varphi 
\, dr \, d\varphi \, dt \qquad .
\end{equation}
Using the fact that $\sqrt{\stackrel {(0)} {\bf g}} \stackrel {(0)} {\cal R}$
is equal to $-2\partial_t \omega_\varphi$ up to a $\varphi$-derivative,
we obtain the expected variation under diffeomorphisms:
\begin{equation}
\delta_{D} {\cal B}^{(t)} = 
 \frac \ell {32 \pi G}
\int_{|r|=\bar r} \stackrel {(0)} {\cal R}
 \sqrt{\stackrel {(0)} {\bf g}} \, \delta \alpha
 \, d\varphi dt
\qquad .
\end{equation}

Let us now consider the variation of the actions under a Weyl transformation
on the spatial boundary $|r|=\bar r$,
which compensates the effect of the diffeomorphism (\ref{difG}) on
the boundary metric:
\begin{equation}
\delta _{W}\stackrel {(0)} g_{ab}= - \delta _{D}\stackrel {(0)} g_{ab}
\qquad .
\end{equation}
The EH action is known to be invariant under the combined action of this 
Weyl transformation and of the diffeomorphism (\ref{difG}) \cite{NB,BERS}:
\begin{equation}
\delta _{D} S_{EH} + \delta _{W} S_{EH} = 0 \qquad .
\end{equation}
The variation of the Liouville action $S_L$ (\ref{SL}) under this Weyl 
transformation is
easily computed using (\ref{Tab2}):
\begin{equation}
\delta_W S_{L}= - \frac \ell {32 \pi G}
\int_{|r|=\bar r} T_{ab} \delta_W \stackrel {(0)} g \strut^{ab}
\sqrt{\stackrel {(0)} {\bf g}} 
 \, d\varphi dt
= - \frac \ell {16 \pi G}\int_{|r|=\bar r}
\stackrel {(0)} {\cal R}
 \sqrt{\stackrel {(0)} {\bf g}} \, \delta \alpha
 \, d\varphi dt
\qquad . 
\end{equation}
Inspection of eqs (\ref{SEHSCS}, \ref{SCSSL}) 
shows that the only other term which 
has a non-vanishing contribution to the Weyl
variation is ${\cal B}_{CS}^{(r)}$.
We find:
\begin{equation}
\delta_W {\cal B}_{CS}^{(r)} = \frac \ell {16 \pi G} \int_{|r|=\bar r}
\frac {4 \bar r ^2} {\ell ^4} \sqrt{\stackrel {(0)} {\bf g}} \, \delta \alpha
 \, d\varphi dt
\end{equation}
We have thus:
\begin{equation}
\delta_W S_{EH}=\delta_W S_L + \delta_W {\cal B}_{CS}^{(r)} \qquad .
\end{equation}

\section{Conclusion}
In this paper we have explicitly shown that the 3-d EH action with
negative cosmological constant is equivalent to a Liouville action
on the spatial $\bar r \rightarrow \infty$ boundaries, plus terms on the
$t$- and $\varphi$-boundaries. The equations of motions derived
from the $\varphi$-boundary terms, in the case of non-trivial topologies
such as those of BTZ black holes,
relate the two non-connected components constituting spatial infinity.
An originality of our approach resides
in the fact that we encoded the holonomy in multi-valued functions.

Owing to the fact that we considered arbitrary curved metrics on the spatial 
$\bar r$ boundaries,
we gave an explicit demonstration that the variation under
diffeomorphisms of the 3-d EH action is equal to the Weyl anomaly of the 
asymptotic Liouville theory.  

We moreover discussed in detail the asymptotically flat solutions, which
generate upon diffeomorphisms all other solutions. We obtained explicitly 
the link between the subdominant term of the metric on the spatial boundary,
which encode the FG ambiguity, and the Liouville field. This reveals that 
though each regular  boundary metric can be expressed in terms of 
a 3-parameter family of Liouville fields, there are
acceptable Liouville fields that lead to singular metrics.
The Liouville (and CS) theory thus contains much more solutions than the
EH gravity theory. 
Let us furthermore stress that a Liouville field defines a solution of
Einstein equations only if it can be extended to CS connections $\bf A$
and $\bf \tilde A$ whose difference yield driebein fields that are regular
or only present singularities interpretable as coordinate singularities. 
In the latter case, 
the topology of the space on which these CS fields are defined 
will differ from that of AdS.

Note that the developments following eq. (\ref{SPCS}), which lead from
the non-chiral WZW action to the Liouville action (\ref{SL}), 
are classically 
valid, but have to be re-examined in the framework of quantum mechanics.
Indeed, in quantum mechanics, the changes of variables leading to 
eq. (\ref{WZW}) and the subsequent elimination of the $X$ and $Y$ 
variables in terms of $\Phi$
involve functional determinants that have been completely ignored.

Finally, our analysis does not solve the problem 
of the degrees of freedom at the origin 
of the black hole entropy, but leads to formulate the hypothesis 
that these degrees of freedom could possibly be encoded in a generalized
holographic principle involving a multiply connected surface at infinity,
each component having its own independent Liouville field only related
to the others by consistency relations dictated by the holonomies.

\vspace*{0.25cm} \baselineskip=10pt{\small \noindent We are grateful
to F. Englert for numerous enlightening discussions and encouragements.
We also acknowledge
K. Bautier, M. Ba\~nados and M. Henneaux for
fruitful discussions.  M.~R. is Senior Research Associate at the Belgian 
National Fund for Scientific Research. This work was partially supported by the a
F.R.F.C. grant and by IISN-Belgium (convention 4.4505.86).}

\section{Appendix}
In this appendix we recall some 2-d formula that we found useful for the
calculations of the main  text.

$sl(2,\real)$ generators:
\begin{equation}
J_0=\frac 1 2 \left ( \begin{array} {cc} 1 & 0 \\ 0 & -1  \end{array}  
\right ) \quad , \quad
J_1=\frac 1 2 \left ( \begin{array} {cc} 0 & 1 \\ 1 & 0  \end{array}  
\right )  \quad , \quad  
J_2=\frac 1 2 \left ( \begin{array} {cc} 0 & -1 \\ 1 & 0  \end{array}  
\right ) 
\end{equation}

Null frame
\begin{eqnarray} \begin{array}{l}
{\underline \theta}^{\hat p}=\theta^{\hat p}_i\, dx^i\quad,\quad 
{\vec e}_{\hat p}=\partial_{\hat p}=e_{\hat p}^i\partial_{x^i} \qquad 
\hat p,\,\hat q\, \dots  \in \left\{ +,\,-\right\} \quad,\quad 
x^i \in \left\{\varphi,\, t\right\}\\
\theta\,e_{\hat  p}^i=\varepsilon_{\hat p\,\hat q}\varepsilon^{i\,
  j}\theta^{\hat p}_j\qquad \varepsilon_{+\,-}=1\quad ,\quad 
\varepsilon^{t\,\varphi}=1
\end{array} \end{eqnarray}

Metric in the null frame
\begin{eqnarray} \begin{array}{lcl}
ds^2 &=& \eta_{\hat p\,\hat q}{\underline \theta}^{\hat p}\otimes 
 {\underline \theta}^{\hat q}\qquad \hat
p,\,\hat q \in \left\{+,\,-\right\}\\
\partial_s^2&=&\eta^{\hat p\,\hat q}{\vec e}_{\hat p}\otimes 
  {\vec e}_{\hat q}\\
\theta&=&\det \left( \theta^{\hat p}_i \right )=\theta^+_t \theta^-_{\varphi}
-\theta^+_{\varphi}
\theta^-_t \\
 \eta _{\hat p \, \hat q} &=& \left(
\begin{array}{rr} 
0&-1/2\\ -1/2&0
\end{array} \right)
\end{array}\end{eqnarray}

Connection coefficients
\begin{eqnarray} \begin{array}{l}
 d {\underline \theta}^+={\underline \omega} \wedge {\underline \theta}^+
\qquad,\qquad 
d {\underline \theta}^-=- {\underline \omega} \wedge
 {\underline \theta}^- \\
\omega= \omega^-_{\ -}=-\omega^+_{\ +}=\omega_+\theta^+ +  \omega_-\theta^- \\
\varepsilon^{i\, j}\varepsilon_{\hat p\, \hat q}\partial_{x^i}
\theta^{\hat  p}_j = \omega_{\hat q}\, \theta
\end{array}\end{eqnarray}

Gauss curvature
\begin{eqnarray} \begin{array}{l}
 {\cal R}=4[\partial_-(\omega_+)-\partial_+(\omega_-)+ 2\,\omega_+\omega_-]\\
\theta\, {\cal R}=4\, \partial_{x^i}[\theta (e^i_-\omega_+ - e^i_+\omega_- )]\\
\end{array}\end{eqnarray}

Weyl variance
\begin{eqnarray} \begin{array}{l}
 {\underline \theta}^{\hat p}\mapsto e^{-\alpha}\, 
{\underline \theta}^{\hat p}\\
{\vec e}_{\hat p}\mapsto e^{\alpha}\,{\vec e}_{\hat p}\\
\theta \mapsto  e^{-2\,\alpha}\theta\\
{\underline \omega} \mapsto {\underline \omega} - 
\partial_-(\alpha)\,{\underline \theta^-} + \partial_+(\alpha)\,
{\underline \theta^+}\\ 
\omega_\pm \mapsto e^{\alpha}\left[\omega_\pm \pm \partial_\pm (\alpha)\right]
\end{array} \end{eqnarray}

Dalembertian
\begin{eqnarray} \begin{array}{l}
\Box \phi=2 \left [ \partial_+ \partial_- \phi - \omega_+ \partial _- \phi
+\partial_- \partial_+ \phi + \omega_- \partial_+ \phi \right ]
\end{array} \end{eqnarray}

Liouville energy-momentum tensor
\begin{equation}
T_{ab}=  \frac 1 2 \phi _{;a} \phi _{;b} -\phi_{;ab} -\eta_{ab}
(\frac 1 4 \phi_{;c} \phi^{;c} - \phi _{;c}^{;c} - \frac 4 {\ell^2} e^\phi)
\end{equation}

\section{Figure caption}
\noindent {\bf Figure 1}: Penrose diagram of BTZ black hole with $ M>\vert
J/\ell \vert$  and domains of
definition of the radial FG coordinate.

\begin{thebibliography}{999}
\bibitem{DJH} S. Deser, R. Jackiw, and G. 't Hooft, Ann. Phys. (N.Y.) {\bf{152}}
  (1984) 220; S. Deser, R.~Jackiw, {\it ibid} {\bf{153}} (1984) 405
\bibitem{BTZ} M. Ba\~nados, C. Teitelboim, and Z. Zanelli, 
  Phys. Rev. Lett. {\bf{69}} (1992) 1849, hep-th/9204099; 
 M. Ba\~nados, M. Henneaux, C. Teitelboim, 
 and Z. Zanelli, Phys. Rev. {\bf{D48}} (1993) 1506, gr-qc/9302012.
\bibitem{CAR} S. Carlip, {\it Quantum Gravity in 2+1 Dimensions}, Cambridge
  Univ. Press (1998)
\bibitem{BH} J. Brown, M. Henneaux, Comm. Math. Phys. {\bf{104}} (1986) 207
\bibitem{ATW} A. Achucarro, and P.K. Townsend, Phys. Lett. {\bf{B180}} 
 (1986) 89;
  E. Witten, Nucl. Phys. {\bf{B311}} (1988) 46.
\bibitem{CHD} O. Coussaert, M. Henneaux, P. van Driel, Class. Quant. Grav. 
   {\bf{12}} (1995) 2961, gr-qc/9506019
\bibitem{MS} G. Moore and N. Seiberg, Phys. Lett. {\bf{B220}} (1989) 422
\bibitem{EMSS} S. Elitzur, G. Moore, A. Schwimmer, N. Seiberg,
  Nucl. Phys. {\bf{B326}}(1989) 108
\bibitem{FWB} P. Forgacs, A. Wipf, J. Balog, L. Feher and L. O'Raifeartaigh,
Phys. Lett. {\bf{B227}} (1989) 213
\bibitem{GL} C.R. Graham, and J.M. Lee, Adv. in Math. {\bf{87}} (1991) 186
\bibitem{FG} C. Fefferman, and C.R. Graham, Soc. Math. de France, 
Ast\'erisque, hors s\'erie (1985) 95
\bibitem{RS} M. Rooman, Ph. Spindel, Ann. Phys. (Leipzig) {\bf 9} (2000) 162,
hep-th/9911142.
\bibitem{NB} N. Boulanger {\it L'anomalie conforme en dimensions 2 et 4},
M\'emoire de Licence U.L.B. (1999), unpublished
\bibitem{BERS} K. Bautier, F. Englert, M. Rooman, Ph. Spindel, Phys. Lett {\bf
    B 479} (2000) 291, hep-th/0002156
\bibitem{MB} M. Ba\~nados, F. Mendez, Phys. Rev. {\bf{D58}} (1998) 104014,
hep-th/9806065
\bibitem{GH} G. W. Gibbons, S. W. Hawking, Phys. Rev. {\bf{D15}} (1977) 2752
\bibitem{BAN}M.  Banados, {\it Three-dimensional quantum geometry and black
holes}, Presented at 2nd La Plata Meeting on Trends in Theoretical
Physics, Buenos Aires, Brazil, 28 Nov.-4 Dec. 1998, hep-th/9901148
\bibitem{FE} F. Englert, private communication.
\bibitem {HT} M. Henneaux, C. Teitelboim, {\it Quantization of gauge systems},
Princeton University Press (1992)
\bibitem{HMS} M. Henneaux, L. Maoz, A. Schwimmer, Annals Phys. {\bf 282} 
   (2000) 31, hep-th/9910013.
\bibitem{WW}E. Whittaker, G. Watson, {\it Modern Analysis-Fourth
    edition}, Cambridge University Press (1965)
\bibitem{Ma}D. Cangemi, M. Leblanc, R.B. Mann, Phys. Rev. {\bf D 48} (1993)
 3606, gr-qc/9211013
\bibitem{CH} O. Coussaert, M. Henneaux, Phys.Rev.Lett. {\bf 72} (1994) 183,
hep-th/9310194.
\bibitem{ISTY}C. Imbimbo, A. Schwimmer, S. Theisen, S. Yankielowicz,
  Class. Quant. Grav. {\bf 17} (2000) 1129, hep-th/9905046
\bibitem{HS} M. Henningson, K.Skenderis, 
JHEP {\bf 9807} (1998) 023, hep-th/9806087.
\bibitem{B} K. Bautier, "Diffeomorphisms and Weyl Transformations in $AdS_3$
Gravity'', Presented at the Meeting on Quantum Aspects of Gauge Theories,
Supersymmetry and Unification, Paris, France, 1-7 Sep. 1999, hep-th/9910134. 
\end{thebibliography}
\end{document}